\documentclass[useAMS,usenatbib]{mn2e}
\usepackage{graphicx,epsfig}
\usepackage{subfigure}
\usepackage{amssymb}
\usepackage{amsmath}

\newcommand{\lunits}{\ensuremath{\,\rmn{erg\,s^{-1}\,Hz^{-1}}}}  % mono luminosity units (cgs)
\newcommand{\ml}{\,\mbox{\ensuremath{\rmn{M_{\sun}\,yr^{-1}}}}}  % mass-loss rate unit
\newcommand{\lsun}{\,\mbox{\ensuremath{\rmn{L_{\sun}}}}}  % solar luminosities
\newcommand{\msun}{\,\mbox{\ensuremath{\rmn{M_{\sun}}}}}  % solar mass
\newcommand{\h}{\ensuremath{\rmn{^h}}}                           % hour (right ascension)
\newcommand{\m}{\ensuremath{\rmn{^m}}}                           % min (right ascension)
\newcommand{\kms}{\,\mbox{\ensuremath{\rmn{km\,s^{-1}}}}}        % km/s  velocity units
\newcommand{\mum}{\,\mbox{\ensuremath{\rmn{\mu m}}}}             % micro meter - wavelength 
\newcommand{\mjy}{\,\mbox{\ensuremath{\rmn{mJy}}}}               % mJy       flux density 
\newcommand{\mujy}{\,\mbox{\ensuremath{\rmn{\mu Jy}}}}           % micro Jy  flux density 
\newcommand{\jyb}{\,\mbox{\ensuremath{\rmn{Jy/beam}}}}           % Jy/beam  peak intensity; rms 
\newcommand{\mujyb}{\,\mbox{\ensuremath{\rmn{\mu Jy/beam}}}}     % muJy/beam  peak intensity; rms 
\newcommand{\ee}[2]{\mbox{\ensuremath{\times10^{{#1}{#2}}}}}     % x 10^yy
\newcommand{\e}[1]{\mbox{\ensuremath{\times10^{{#1}}}}}          % x 10^y
\newcommand{\eee}[3]{\mbox{\ensuremath{\times10^{{#1}{#2}{#3}}}}} % x 10^yyy
\newcommand{\uv}{\mbox{{\it uv}-}}                                                       % uv- (plane; coverage)
\newcommand{\tpeak}{\ensuremath{t_{\rmn{peak}}}}                                         % Peak time
\newcommand{\lpeak}{\ensuremath{L_{\rmn{peak}}}}                                         % Peak luminosity
\newcommand{\speak}{\ensuremath{S_{\rmn{peak}}}}                                         % Peak intensity
\newcommand{\lir}{\ensuremath{L_{\rmn{IR}}}}                                             % IR luminosity
\newcommand{\alphxc}{\ensuremath{\alpha_{\scriptstyle C}^{\scriptstyle X}}} % spectral index btw nu_x and nu_c
\newcommand{\alphcl}{\ensuremath{\alpha_{\scriptstyle L}^{\scriptstyle C}}} % spectral index btw nu_c and nu_l
\newcommand{\tb}{\ensuremath{T_{\rmn{B}}}}                                               % brightness temperature
\newcommand{\hii}{\mbox{H{\footnotesize{II}}}}                                           % HII region
\title[SNe 2010O and 2010P in Arp\,299 - II]
{The nature of supernovae 2010O and 2010P in Arp\,299 - II. Radio emission}
\author[Romero-Ca\~nizales et al.]
  {C.\ Romero-Ca\~nizales,$^{1,2}$\thanks{E-mail: cromero@astro.puc.cl}
  R.\ Herrero-Illana,$^3$
  M.\ A.\ P\'erez-Torres,$^3$
  A.\ Alberdi,$^3$ 
\newauthor 
  E.\ Kankare,$^{4,2}$
  F.\ E.\ Bauer,$^{1,5}$
  S.\ D.\ Ryder,$^6$
  S.\ Mattila,$^{4}$
  J.\ E.\ Conway,$^7$
 \newauthor 
  R.\ J.\ Beswick$^8$ and 
  T.\ W.\ B.\ Muxlow$^8$\\  
$^1$Instituto de Astrof\'{\i}sica, Facultad de F\'{\i}sica, Pontificia Universidad 
Cat\'olica de Chile, Casilla 306, Santiago 22, Chile \\
$^2$Tuorla Observatory, Department of Physics and Astronomy, University of Turku, V\"ais\"al\"antie 20,  
    FI-21500 Piikki\"o, Finland \\
$^3$Instituto de Astrof\'{\i}sica de Andaluc\'{\i}a -- CSIC, PO Box 3004, 18080 Granada, Spain \\
$^4$Finnish Centre for Astronomy with ESO (FINCA), University of Turku, V\"ais\"al\"antie 20, FI-21500 
    Piikki\"o, Finland \\
$^5$Space Science Institute, 4750 Walnut Street, Suite 205, Boulder, CO 80301, USA \\
$^6$Australian Astronomical Observatory, PO Box 915, North Ryde, NSW 1670, Australia \\
$^7$Onsala Space Observatory, SE-439 92 Onsala, Sweden  \\
$^8$Jodrell Bank Centre for Astrophysics, The University of Manchester, Oxford Rd, Manchester M13 9PL, UK
}

\begin{document}

\date{Accepted 2014 March 4.  Received 2014 March 4; in original form 2013 October 14}
\pagerange{\pageref{firstpage}--\pageref{lastpage}} \pubyear{2014}

\maketitle

\label{firstpage} 

\begin{abstract} 
We report radio observations of two stripped-envelope supernovae (SNe), 2010O and 2010P, which 
exploded within a few days of each other in the luminous infrared galaxy Arp\,299. Whilst SN\,2010O 
remains undetected at radio frequencies, SN\,2010P was detected (with an astrometric accuracy better 
than 1\,milli arcsec in position) in its optically thin phase in epochs ranging from $\sim1$ to 
$\sim3$\,yr after its explosion date, indicating a very slow radio evolution and a strong interaction 
of the SN ejecta with the circumstellar medium. Our late-time radio observations toward SN\,2010P probe 
the dense circumstellar envelope of this SN, and imply $\dot M [\ml{}] / v_{\rmn{wind}} [10\,\kms{}]
=(3.0$--5.1)$\ee-5$, with a 5\,GHz peak luminosity of $\sim1.2\ee27$\,\lunits{} on day $\sim$464 after explosion. 
This is consistent with a Type IIb classification for SN\,2010P, making it the most distant and most slowly 
evolving Type IIb radio SN detected to date.
\end{abstract}


\section[]{Introduction}\label{sec:sneinarp299}

Core-collapse supernovae (CCSNe) are the signposts of recent massive star formation. Since much of the 
massive star formation is embedded in dust, a substantial fraction of the CCSNe in the Universe will 
remain hidden in optical searches \citep{mattila12}. This is particularly true in the case of star 
formation taking place in the dusty environments of luminous ($10^{11}\,\lsun{} < \lir{} < 
10^{12}\,\lsun{}$)\footnote{We adopt $\lir = L[8$--$1000\mum{}]$.}, and ultra-luminous ($\lir{} > 
10^{12}\,\lsun{}$) infrared (IR) galaxies (LIRGs and ULIRGs, respectively), which dominates the star 
formation rate density at $z>1$ \citep[e.g.,][]{magnelli11}. In many LIRGs, the bulk of star formation 
occurs within their circumnuclear regions ($<1$\,kpc), so that the need for high-resolution observations 
becomes crucial in the detection and study of CCSNe therein.

The synergy between high resolution radio and near-infrared (NIR) {\it K}-band (where the extinction is 10 times 
lower than in the optical) observations is currently being used to help build a complete picture of the supernova 
(SN) activity in dusty starbursts and LIRGs. Outstanding examples are SNe 2000ft \citep[in NGC\,7469;][and references 
therein]{perez09a}, 2004ip \citep[in IRAS\,18293$-$3413;][]{mattila07,perez07}, 2008cs \citep[in IRAS\,17138$-$1017;][]{kankare08}, 
and 2008iz \citep[in M82;][]{brunthaler09, mattila13} where both high resolution (to disentangle the emission of the SN from its host) 
and reduced extinction measurements (radio and NIR), were essential in their detection and characterisation.

Arp\,299 is a LIRG with an IR luminosity ($\lir \approx6.7\ee11$\,L$_{\sun}$ 
\citep{sanders03}, at an adopted luminosity distance of 44.8\,Mpc, assuming $H_0 = 73$\,km\,s$^{-1}$\,Mpc$^{-1}$.
The system is composed of two interacting galaxies, whose major nuclei (A and B1) and core components in the 
interacting region (C$^{\prime}$ and C) are bright radio and NIR emitters \citep[][]{gehrz83, alonso00}. Arp\,299 
is a very prolific SN factory as proved by the detection of several radio SNe and SN remnants (SNRs) in the innermost 
nuclear regions of Arp\,299A and Arp\,299B \citep{perez09b, ulvestad09, rocc11, bondi12}, and the detection within 
the last 20 years of seven optical/NIR SNe in the circumnuclear regions of the system \citep[see][]{anderson11, 
mattila12}. 

Many attempts have been made to detect at radio wavelengths the SNe occurring in the circumnuclear regions of 
Arp\,299, mostly under observing programmes with the Very Large Array (VLA): AS333 carried out from 1990 May 
to 1993 December; AS525 on 1994 February to detect SN\,1993G; AS568 on 1999 January, February, April and October 
to detect SN\,1999D; AW641 on 2005 February, June and August to detect SN\,2005U. We are currently monitoring 
Arp\,299 at radio wavelengths under programmes AP592 and AP614, ``Unveiling the Hidden Population of SNe in Local 
Luminous Infrared Galaxies'' (PI: M.\ A.\ P\'erez-Torres), as part of a combined radio and NIR SN search in a sample 
of local LIRGs. In this paper we study the nature of the most recently detected SNe in the system, 2010O and 2010P, 
by means of their late-time radio emission. 

SN\,2010O was discovered at optical wavelengths on 2010 January 24 \citep{newton10}. The object exploded on 2010 
January 7 and lies on a location with intermediate extinction, $A_V =2$\,mag (Kankare et al. 2013). It was classified 
as a Type Ib SN by \citet{mattila10b} based on a low-resolution optical spectrum obtained with the Nordic Optical 
Telescope (NOT). \citet{nelemans10} reported an X-ray transient at the position of SN\,2010O prior to its explosion, 
and suggested that the progenitor of this SN was part of a Wolf-Rayet X-ray binary system, similar to those found in 
our Galaxy.

SN\,2010P was discovered at NIR wavelengths with the NOT on 2010 January 18 \citep{mattila10}, a few days after 
explosion (2010 January 10, Kankare et al. 2013). The spectrum of SN\,2010P obtained with the Gemini-North Telescope 
revealed a deficiency of hydrogen and matched with spectra of Type Ib/IIb SNe \citep{ryder10}, and the absolute 
magnitude and colours from optical and NIR observations with the NOT and the Gemini-North Telescope yielded a 
likely high host galaxy extinction of $A_V \sim 7$ (Kankare et al. 2013).

This paper accompanies \citeauthor{kankare13} by Kankare et al. (2013) which presents a study of the early 
optical and NIR emission of SNe 2010O and 2010P. Here we analyse long-term follow-up radio observations of 
Arp\,299 carried out by us and obtained from archives, with particular emphasis on characterising SNe 2010O 
and 2010P. In Section \ref{sec:radio} we give details on the data reduction and analysis. We describe our 
results in Section \ref{sec:results}, which we then discuss in Section \ref{sec:discussion}. In Section 
\ref{sec:concl} we summarize the conclusions of our work. 

\section[]{Radio observations}
\label{sec:radio}

We have collected observations of Arp\,299 obtained with the Multi-Element Radio Linked Interferometry Network 
(MERLIN), the electronic Multi-Element Remotely Linked Interferometer Network (e-MERLIN\footnote{e-MERLIN is the 
UK's facility for high resolution radio astronomy observations, operated by The University of Manchester for the 
Science and Technology Facilities Council.}), the VLA of the National Radio Astronomy Observatory (NRAO\footnote{NRAO 
is a facility of the National Science Foundation operated under cooperative agreement by Associated Universities, 
Inc.}), and the European very long baseline interferometry (VLBI) Network (EVN\footnote{The European VLBI Network is 
a joint facility of European, Chinese, South African and other radio astronomy institutes funded by their national 
research councils.}). In Table \ref{tab:radobs} we show basic information for these observations (all of which have 
an assigned label) including the range of frequencies observed, the total time on target ($t_{\rmn{on}}$), and
the peak intensity in each epoch of J1128$+$5925, which was used as the phase calibrator in all the observations we 
report here. 

%%%%%%%%%%%%%%%%%%%%%%%%%%%%%%%%%%%%%%%%%%%%%%%%%%%%%%%%%%%%%%%%%%%%%%%%%%%%%%%%%%%%%%%%%%%%%%%%%%%%%%%%%%%%%%%%%% 
\begin{table*}
\centering 
\caption{\protect{Arp\,299 radio observations.}} \label{tab:radobs}
\begin{tabular}{cllclcc} \hline
\multicolumn{1}{c}{Label} &  \multicolumn{1}{c}{Project} & \multicolumn{1}{c}{Observing}& \multicolumn{1}{c}{ Frequency range}  &
\multicolumn{1}{c}{Array / Participating stations} & \multicolumn{1}{c}{$t_{\rmn{on}}$} & \multicolumn{1}{c}{$P_{\nu}$[J1128$+$5925]}  \\
 & & \multicolumn{1}{c}{date} & \multicolumn{1}{c}{(MHz)} & \multicolumn{1}{c}{} & ~ & \multicolumn{1}{c}{(\jyb{})} \\
\hline
E1 & -      & 2010-01-29/02-01 & 4986--5002   & MERLIN / Mk2, Kn, De, Pi, Da, Cm                & 29\,h & 0.44 $\pm$ 0.02 \\
E2 & AL746  & 2011-03-29       & 28500--29500 & VLA - B configuration                           & 6\,m  & 0.31 $\pm$ 0.02 \\
E3 & AL746  & 2011-03-29       & 35500--36500 & VLA - B configuration                           & 6\,m  & 0.28 $\pm$ 0.01 \\
E4 & AP592  & 2011-06-15/06-18 & 8372--8628   & VLA - A configuration                           & 19\,m & 0.44 $\pm$ 0.05 \\
E5 & -      & 2011-07-04       & 4444--4956   & e-MERLIN / Mk2, Kn, De, Pi, Da                        & 25\,h & 0.40 $\pm$ 0.10 \\
E6 & EP075B & 2012-05-27       & 8344--8472   & EVN / Ef, Wb, On, Mc, Nt, Ur, Ys, Sv, Zc              & 4.3\,h& 0.42 $\pm$ 0.02 \\
E7 & EP075C & 2012-06-04       & 4919--5047   & EVN / Ef, Wb, Jb1, On, Mc, Nt, Tr, Ur, Ys, Sv, Zc, Bd & 2.7\,h& 0.50 $\pm$ 0.03 \\
E8 & EP075D & 2012-06-14       & 1587--1715   & EVN / Ef, Wb, Jb1, On, Mc, Nt, Tr, Ur, Sv, Zc, Bd     & 2.9\,h& 0.36 $\pm$ 0.02 \\
E9 & AP614  & 2012-10-20       & 7395--9395   & VLA - A configuration                           & 19\,m & 0.62 $\pm$ 0.03 \\
E10& EP075E & 2012-10-31       & 4919--5047   & EVN / Ef, Wb, Jb2, On, Mc, Tr, Ur, Ys, Bd, Sh   & 4.4\,h& 0.58 $\pm$ 0.03 \\
E11& AP614  & 2012-11-05       & 7395--9395   & VLA - A configuration                           & 19\,m & 0.60 $\pm$ 0.03 \\
E12& AP614  & 2012-11-21       & 7395--9395   & VLA - A configuration                           & 19\,m & 0.62 $\pm$ 0.03 \\
E13& AP614  & 2012-12-26       & 7395--9395   & VLA - A configuration                           & 19\,m & 0.59 $\pm$ 0.03 \\
    \hline
   \end{tabular}  
\begin{flushleft}
\vspace{-1mm}
Last column: Peak intensity of the phase reference source.
\end{flushleft}
\end{table*}
%%%%%%%%%%%%%%%%%%%%%%%%%%%%%%%%%%%%%%%%%%%%%%%%%%%%%%%%%%%%%%%%%%%%%%%%%%%%%%%%%%%%%%%%%%%%%%%%%%%%%%%%%%%%%%%%%% 

The reduction and analysis for MERLIN, e-MERLIN and EVN data were made following standard procedures within 
the NRAO Astronomical Image Processing System ({\sc aips}); for the VLA data, we used the Common Astronomy Software 
Applications package \citep[{\sc casa};][]{mcmullin07}.

Epoch E1 with MERLIN (PI: M.\ A.\  P\'erez-Torres) was obtained in an attempt to detect the early radio 
emission of SNe 2010O and 2010P \citep[results reported in][]{beswick10}. Observations of 3C\,286 and OQ208 
were used for absolute flux density and bandpass calibration.

We used publicly available Ka-band VLA data taken in B configuration (maximum antenna separation 
of $\approx 11$\,km) under project AL746 (PI: A.\ K.\ Leroy), which include two sub-bands at 29.0 and 
36.0\,GHz, which we labelled for convenience E2 and E3, respectively, although these observations were 
obtained at the same epoch. 3C\,286 was observed for absolute flux density calibration. To obtain the 
final image, we applied standard imaging procedures and two iterations of phase-only self-calibration 
on the target. Since we were interested in a possible detection of the SN, we applied natural weighting 
to the data to minimise the root mean square (rms).

We observed Arp\,299 with the VLA in its most extended configuration, A ($\approx 36$\,km 
as the maximum antenna separation), under program AP592, and labelled 
here as epoch E4 \citep[the first radio detection of SN\,2010P;][]{herrero12}. For imaging we followed
the same procedure as for epochs E2 and E3, except for the use of uniform weighting instead of natural, 
to reduce the effect of confusion with the extended emission (traced by the shortest baselines) 
at the SN site. This is also true for epochs E9, E11, E12 and E13 which were obtained under
program AP614.

Epoch E5 was observed as part of the commissioning observations of e-MERLIN for the Luminous 
Infrared Galaxy Inventory legacy project (LIRGI;\footnote{http://lirgi.iaa.es} PIs: J., Conway \& 
M.\ A.\ P\'erez-Torres). The total on-source time was $\sim25$\,hr (see Table \ref{tab:radobs}) but 
intense data flagging was needed, leaving only $\sim16$\,hr worth of useful data. As amplitude 
calibrators we observed 3C\,286, OQ208 and DA193; to correct the bandpass we used DA193 and 1803$+$784, 
the last one being used also as a fringe finder.

Our EVN observations (epochs E6, E7, E8 and E10) originally were aimed at imaging the Arp\,299A 
and -B1 nuclear regions. After confirming the radio detection of SN\,2010P in epoch E4 with the 
VLA \citep[see][]{herrero12}, we requested the correlation of the EVN observations at the position 
of the radio SN. For all these epochs, the source 4C39.25 was used as a fringe finder, and also 
together with J1128$+$5925 as a bandpass calibrator, to ensure having solutions for all the 
antennas. We used the EVN pipeline products and improved the calibration of the data by removing 
radio interference artifacts and by including ionospheric corrections. We determined gain 
corrections for each antenna by imaging the calibrators with the Caltech program {\sc difmap} 
\citep{difmap}, and applied corrections larger than 10 per cent to the \uv{}data using the 
{\sc aips} task {\sc clcor}. The EVN maps of SN\,2010P at the different frequencies were all 
made with natural weighting.

In Table \ref{tab:measurements} we show the characteristics of the resultant images, such as the 
synthesised beam, the attained off-source rms noise, together with other relevant 
information that will be described in the following sections. 

\section[]{Results}\label{sec:results}

SNe 2010O and 2010P were not detected in the MERLIN observations carried out between 2010 January 29 
and February 1 at 5.0\,GHz \citep[][]{beswick10}. However, late-time radio observations have shed light 
on the nature of these SNe.

%%%%%%%%%%%%%%%%%%%%%%%%%%%%%%%%%%%%%%%%%%%%%%%%%%%%%%%%%%%%%%%%%%%%%%%%%%%%%
\begin{table*}
\centering
\caption{\protect{Observational data measured from the radio images of SN\,2010P obtained from the observations 
described in Table \ref{tab:radobs}.}} 
\label{tab:measurements}
\begin{tabular}{cccccccc} \hline
Label & $t_{\rmn{obs}}-t_0$ & $\nu$  &  rms  & FWHM, PA & $S_{\nu}$ &  $L_{\nu}$ & log\,\tb{} \\
 & (days) & (GHz) & (\mujyb{}) &  (mas$^2$, $\degr{}$) & (\mujy{})  & ($10^{26}$\lunits{}) & (K) \\
(1) & (2) & (3) & (4) & (5) & (6) & (7) & (8) \\
\hline
E1 &   19 & 5.0      &  62  & $46.4\times44.1$, $-75.1$  &  $< 186$       &  -             &   -             \\
E2 &  443 & 29.0     &  47  & $304\times285$, 86.6       &  278 $\pm$ 49  &  6.7 $\pm$ 1.2 & $<$0.67 $\pm$ 0.08 \\
E3 &  443 & 36.0     &  83  & $305\times260$, $-69.4$    &  $< 249$       &  -             &   -             \\
E4 &  521 & 8.5      &  88  & $211\times151$, 12.7       &  541 $\pm$ 92  & 13.0 $\pm$ 2.2 & $<$2.46 $\pm$ 0.07 \\
E5 &  540 & 4.7      &  63  & $187\times143$, 47.5       &  585 $\pm$ 69  & 14.0 $\pm$ 1.7 & $<$3.09 $\pm$ 0.05 \\
E6 &  868 & 8.4      &  18  & $2.7\times1.5$, 21.0       &  123 $\pm$ 19  &  3.0 $\pm$ 0.5 & $<$6.26 $\pm$ 0.07 \\
E7 &  876 & 5.0      &  17  & $4.2\times3.1$, 34.6       &  288 $\pm$ 22  &  6.9 $\pm$ 0.5 & $<$6.04 $\pm$ 0.03 \\
E8 &  886 & 1.7      &  15  & $17.3\times8.1$,7.9        &  466 $\pm$ 28  & 11.2 $\pm$ 0.7 & $<$6.17 $\pm$ 0.03 \\
E9 & 1014 & 8.5      &  51  & $226\times135$, 59.7       &  259 $\pm$ 52  &  6.2 $\pm$ 1.3 & $<$2.16 $\pm$ 0.09 \\
E10& 1025 & 5.0      &  13  &$6.2\times3.9$,$-78.0$      &  284 $\pm$ 19  &  6.8 $\pm$ 0.5 & $<$5.76 $\pm$ 0.03 \\
E11& 1030 & 8.5      &  46  &$188\times136$,$-177.2$     &  270 $\pm$ 48  &  6.5 $\pm$ 1.2 & $<$2.26 $\pm$ 0.08 \\
E12& 1046 & 8.5      &  41  & $183\times138$, 8.7        &  195 $\pm$ 42  &  4.7 $\pm$ 1.0 & $<$2.12 $\pm$ 0.09 \\
E13& 1081 & 8.5      &  43  & $225\times133$, 62.6       &  251 $\pm$ 45  &  6.0 $\pm$ 1.1 & $<$2.15 $\pm$ 0.08 \\
    \hline
   \end{tabular} 
\begin{flushleft}
{\it Columns}: (1) Epoch label.  (2) Days since explosion. (3) Observing central frequency. (4) rms noise. (5) Full width 
at half maximum (FWHM) synthesised interferometric beam. (6) Flux density. (7) Monochromatic luminosity at the observing 
frequency.  (8) Brightness temperature. To calculate the solid angle subtended by the SN we used the major and minor axes 
from column 5, except for epoch E6 where we could obtain the deconvolved size by means of a Gaussian fitting with the 
{\sc aips} task {\sc imfit}.
\end{flushleft}
\end{table*}
%%%%%%%%%%%%%%%%%%%%%%%%%%%%%%%%%%%%%%%%%%%%%%%%%%%%%%%%%%%%%%%%%%%%%%%%%%%%%%%%%%

\subsection[]{The radio non-detection of SN\,2010O}\label{ssec:sn2010o}

In \citeauthor{kankare13} we determined that SN\,2010O exploded on 2010 January 7. We do not detect it in either early- 
($\sim0.6$\,month) or late-time ($\sim1$ to $\sim3$\,yr after the explosion) radio observations reported by \citet{beswick10} 
and in this paper, respectively. The MERLIN observations (E1 at $\sim0.6$\,month) by \citeauthor{beswick10} resulted in a 
$3\sigma$ upper limit of 186\mujyb{} for the radio emission of SN\,2010O, i.e., a luminosity $<4.5\ee26$\,\lunits{}. The most 
sensitive late-time epoch we present here which covers the SN\,2010O region (E12 at 2.8\,yr), results in a $3\sigma$ upper 
limit of 123\mujyb{} ($<3\ee26$\,\lunits{}). Our results suggest that: {\it i)} the SN was intrinsically weak, with a luminosity 
$<3\ee26$\,\lunits{}; and/or {\it ii)} its radio emission was heavily absorbed by the dense CSM at the moment of the early-time 
MERLIN observations, and owing to a fast-evolving nature, its short radio lifetime prevented its detection in our late-time radio
observations. 

\subsection[]{The radio emission of SN\,2010P}\label{ssec:radem}

In \citet{herrero12} we reported the identification of SN\,2010P with a new high signal-to-noise ratio (S/N$>10$) 
radio source in our 8.5\,GHz VLA map from 2011 June 15-18 (E4), matching the NIR position within its 1$\sigma$ 
uncertainty reported by \citet[][]{mattila10}. No radio source is visible at that position in any of the 14 VLA 
epochs from 1990 to 2006 analysed by \citet{rocc11}. The radio detection of SN\,2010P is confirmed by our e-MERLIN 
(4.7\,GHz) and EVN observations (at 1.7, 5.0 and 8.4\,GHz), and by the archival VLA data, all of these
made during 2011 and 2012. 

%%%%%%%%%%%%%%%%%%%%%%%%%%%%%%%%%%%%%%%%%%%%%%%%%%%%%%%%%%%%%%%%%%%%%%%%%%%%%%%%%%
\begin{figure*}
\centering
\includegraphics[width=168mm]{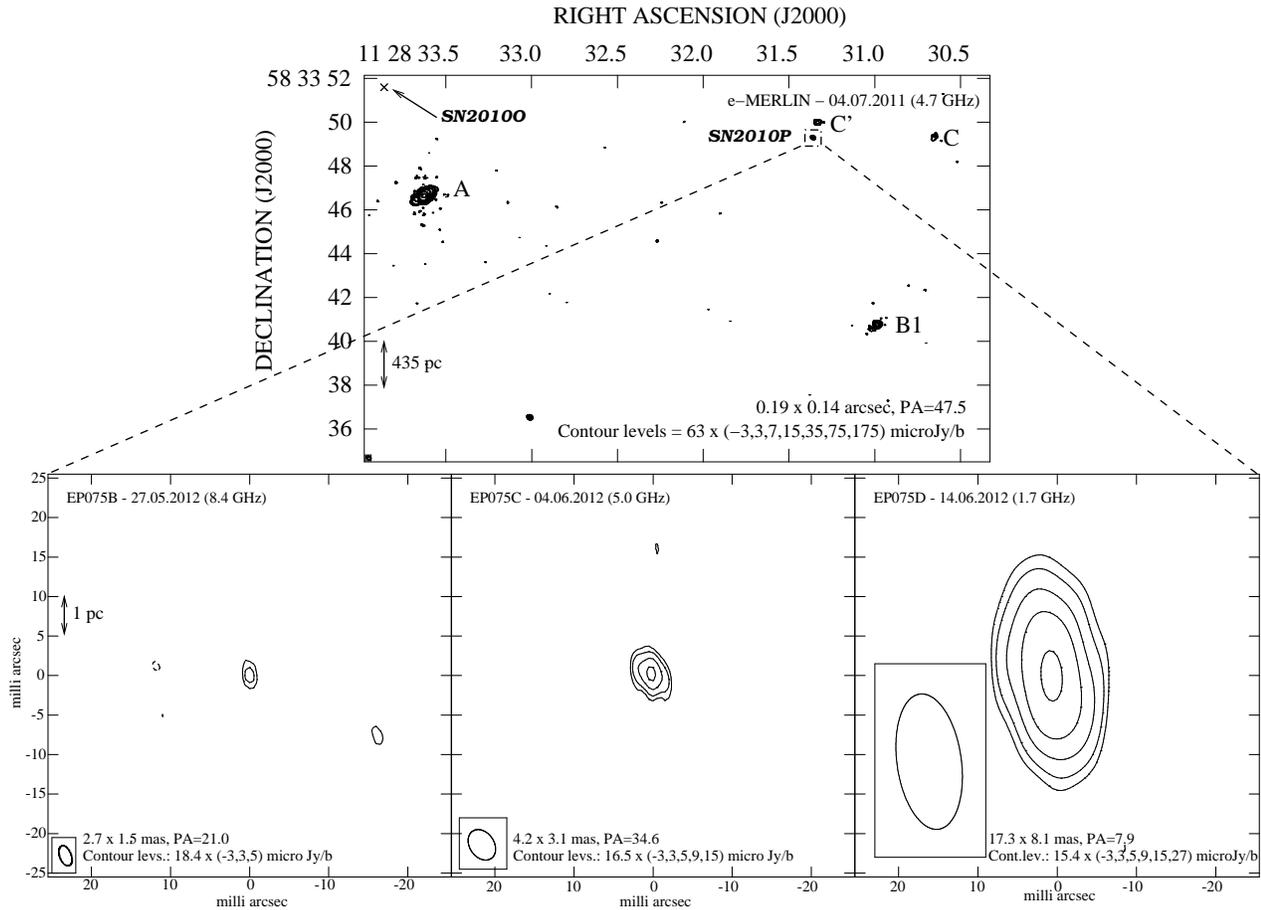}
\caption{Top: e-MERLIN contour image of Arp\,299 at a median central frequency of 4.7\,GHz with an rms noise of
         63\,\mujyb{}. The main radio sites are indicated with the letters A, B, C$^{\prime}$ and C, the position of 
         SN\,2010O is represented by a cross, and SN\,2010P is enclosed by a box. Bottom: EVN contour images of SN\,2010P 
         at a median central frequency of 8.4, 5.0 and 1.7\,GHz (from left to right). All the EVN maps are centred 
         at $\alpha(J2000) = 11\h28\m31\fs3605$, $\delta(J2000) = 58\degr33\arcmin49\farcs315$. The rms noise in each 
         EVN image is 18, 17 and 15\,\mujyb{}, respectively. The beam size is shown in the lower left corner of each 
         contour image.} 
\label{fig:arp299}
\end{figure*}
%%%%%%%%%%%%%%%%%%%%%%%%%%%%%%%%%%%%%%%%%%%%%%%%%%%%%%%%%%%%%%%%%%%%%%%%%%%%%%%%%%

In Table \ref{tab:measurements} we report the parameters measured from our radio maps at the position of SN\,2010P 
(e-MERLIN and EVN maps are shown in Figure \ref{fig:arp299}). We include the convolving beam, as well as the attained 
rms and the flux density measurements in each map. Since the SN remains unresolved in all the epochs, we take its 
measured peak intensity as a good approximation for the flux density, whose uncertainty includes the contributions of 
the rms and a systematic 5 per cent uncertainty in the absolute flux calibration\footnote{We assume this conservative 
systematic error at all frequencies and for measurements with all the different arrays included here, although the error 
might be lower. For instance, in the radio observations of SN\,2011dh with the VLA, a 1 per cent systematic error was 
considered for $\nu<20$\,GHz, and only a 3 per cent error for $\nu<20$\,GHz \citep{krauss12}.}. For the images presented 
in Figure \ref{fig:arp299} we modified the pixel size of the maps (from epochs E6, E7 and E8) to match that of the lowest 
resolution map obtained at 1.7\,GHz, using the task {\sc ohgeo} within {\sc aips}.

The luminosities obtained in the late time observations are typical of normal CCSNe. The fact that the SN is detected 
$\sim1$\,yr after explosion and that it has remained radio bright for about two years (the time between epochs E2 and E13), 
indicates a slow radio evolution and thus the presence of either a dense and/or extended circumstellar medium (CSM).

At an age of 2.4 and 2.8\,yr (epochs E6 to E8 and E10) the $\tb$ values indicate that the radio emission is 
non-thermal at the three different frequencies observed with the EVN, and corresponds to a compact object
of 2.0 $\times$ 0.6 mas$^2$ (deconvolved size) at $\alpha(J2000) = 11\h28\m31\fs3605 \pm 0.2$\,mas, 
$\delta(J2000) = 58\degr33\arcmin49\farcs315 \pm 0.2$\,mas in our 8.4\,GHz EVN map (E6), which provides the highest 
resolution and thus the highest astrometric accuracy. The uncertainty in position has been determined by adding in
quadrature $\rmn{FWHM}/(2 \times \rmn{S/N})$ of the SN, plus the uncertainty in position of the phase reference
source ($\Delta\alpha=0.23$\,mas, $\Delta\delta=0.13$\,mas)\footnote{L. Petrov, solution rfc\_2012b (unpublished, 
available on the Web at http://asrtogeo.org/vlbi/solutions/rfc\_2012b).}.

We conservatively take the deconvolved size from epoch E6 to set an upper limit to the size of the expanding ejecta 
of 1.3\,mas in diameter on average, which at a distance of 44.8\,Mpc, implies an upper limit of 4.3\ee17\,cm in radius. 
For any SN progenitor, such a size would imply that the distance from either the progenitor's centre or the photosphere 
to the shock region are basically equal. Thus, even considering free-expansion, it is not possible to set a reliable 
constraint for the shock velocity.

The $\tb$ values from the VLA and e-MERLIN epochs are consistent with non-thermal emission 
that is being suppressed owing to thermal and/or non-thermal processes. With a larger beam
in those epochs (compared to that from our EVN images) it is likely that the flux of the SN 
itself and the contaminating emission from the host, will contribute jointly to the flux 
measurements. 

\subsubsection[]{Contaminating emission}

A meaningful comparison among epochs with different resolution necessitates an estimate of the background 
emission affecting the flux density measurements of the low-resolution images. Contamination is likely
to occur when observing at angular resolutions comparable to, or only a few times the separation between a SN 
and its host. This is specially important in this case since SN\,2010P exploded very close to component C$^{\prime}$ 
($\approx 160$\,pc projected distance, or $\approx 720$\,mas), which is also a bright radio source and  
is surrounded by diffuse emission.

The diffuse emission detectable in the different low-resolution epochs (e.g., E5, Figure \ref{fig:arp299})
is limited by the thermal rms in the individual images and the weighting used in the mapping process. The 
wealth of diffuse emission at the position of SN\,2010P is revealed in deep Arp\,299 images at different 
frequencies \citep[e.g.,][]{neff04}.

We have co-added all the 8.5\,GHz epochs analysed by \citet{rocc11}, and produced a deep image with 
a convolving beam of $0.21\times0.14$\,arcsec$^2$, equal to the average of the convolving beam used in our 
VLA epochs at 8.5\,GHz. We then solved for the zero level emission at the position of SN\,2010P 
using the task {\sc imfit} within {\sc aips}, which we then adopted as the background emission at 
8.5\,GHz for SN\,2010P. 

We estimate the background emission at 4.7\,GHz by applying the fiducial spectral index obtained by 
\citet{leroy11} for Arp\,299 ($\alpha = -0.62$), to the background emission at 8.5\,GHz. Such a spectral
index does not apply to frequencies $\ga30$\,GHz where a higher contribution from thermal emission is
expected. Instead, we obtained a representative zero level emission at 29.0\,GHz, by measuring it
at different positions around the SN location in our E2 image, and then taking the average of the 
retrieved values. We note that the E2 image was made with natural weighting to facilitate a robust SN 
detection, and thus some diffuse emission was detectable in it. The background emission at 4.7, 8.5 and 
29.0\,GHz are given in Table \ref{tab:bge}.

We subtracted the background emission from epochs E2, E4, E5, E9, E11, E12 and E13, at their corresponding
frequencies. We then calculated the E4 to E5 and the E6 to E7 spectral indices, to convert the e-MERLIN 
flux density to be at 5.0\,GHz, and allow its comparison with epochs E1 and E7. Similarly, we converted 
the E6 flux density to be at 8.5\,GHz, for the sake of homogeneity. The corrected flux densities at 
the adopted frequencies along with the observed parameters (from Table \ref{tab:radobs}) are shown in Table 
\ref{tab:fluxcorr} for comparison.

%%%%%%%%%%%%%%%%%%%%%%%%%%%%%%%%%%%%%%%%%%%%%%%%%%%%%%%%%%%%%%%%%%%%%%%%%%%%%%%%%%
\begin{center}
\begin{table}
\centering
\caption{\protect{Contaminating background emission for SN\,2010P.}} \label{tab:bge}
\begin{tabular} {cc} \hline
\multicolumn{1}{c}{Frequency} & \multicolumn{1}{c}{$S_{\nu}$} \\
\multicolumn{1}{c}{(GHz)} &  \multicolumn{1}{c}{(\mujy{})} \\
    \hline
4.7  & 178 $\pm$ 10 \\
8.5  & 123 $\pm$  7 \\
29.0 & 200 $\pm$ 23 \\
    \hline 
   \end{tabular}
\end{table}
\end{center}
%%%%%%%%%%%%%%%%%%%%%%%%%%%%%%%%%%%%%%%%%%%%%%%%%%%%%%%%%%%%%%%%%%%%%%%%%%%%%%%%%%

%%%%%%%%%%%%%%%%%%%%%%%%%%%%%%%%%%%%%%%%%%%%%%%%%%%%%%%%%%%%%%%%%%%%%%%%%%%%%%%%%%
\begin{table*}
\centering  
\caption{\protect{ SN\,2010P measured and contamination-free flux densities.  }} \label{tab:fluxcorr}
\begin{tabular} {cccccc} \hline
\multicolumn{2}{c}{Epoch}  & \multicolumn{2}{c}{Observed} &\multicolumn{1}{c}{Adopted} & \multicolumn{1}{c}{Corrected} \\
\multicolumn{1}{c}{label} & \multicolumn{1}{c}{$t_{\rmn{obs}}-t_0$} & \multicolumn{1}{c}{$\nu$} & 
               \multicolumn{1}{c}{$S_{\nu}$} & \multicolumn{1}{c}{$\nu$} & \multicolumn{1}{c}{$S_{\nu}$} \\
~ & \multicolumn{1}{c}{(days)} &\multicolumn{1}{c}{(GHz)} &  \multicolumn{1}{c}{(\mujy{})}
&\multicolumn{1}{c}{(GHz)} &  \multicolumn{1}{c}{(\mujy{})} \\
    \hline
E1  &   19 & 5.0   & $< 186$       &  5.0 &  $< 186$     \\
E2  &  443 & 29.0  & 278 $\pm$ 49  & 29.0 &  76 $\pm$ 72 \\
E3  &  443 & 36.0  & $< 249$       & 36.0 &  $< 249$     \\
E4  &  521 & 8.5   & 541 $\pm$ 92  & 8.5  & 418 $\pm$ 99 \\
E5  &  540 & 4.7   & 585 $\pm$ 69  & 5.0  & 408 $\pm$ 81 \\
E6  &  868 & 8.4   & 123 $\pm$ 19  & 8.5  & 122 $\pm$ 19 \\
E7  &  876 & 5.0   & 288 $\pm$ 22  & 5.0  & 288 $\pm$ 22 \\
E8  &  886 & 1.7   & 466 $\pm$ 28  & 1.7  & 466 $\pm$ 28 \\
E9  & 1014 & 8.5   & 259 $\pm$ 52  & 8.5  & 136 $\pm$ 59 \\
E10 & 1025 & 5.0   & 284 $\pm$ 19  & 5.0  & 284 $\pm$ 19 \\
E11 & 1030 & 8.5   & 270 $\pm$ 48  & 8.5  & 147 $\pm$ 55 \\
E12 & 1046 & 8.5   & 195 $\pm$ 42  & 8.5  &  71 $\pm$ 49 \\
E13 & 1081 & 8.5   & 251 $\pm$ 45  & 8.5  & 127 $\pm$ 52 \\
    \hline 
   \end{tabular}
\begin{flushleft}
Note that the values of some epochs are unchanged, but we show them all for completeness. 
\end{flushleft}
\end{table*}
%%%%%%%%%%%%%%%%%%%%%%%%%%%%%%%%%%%%%%%%%%%%%%%%%%%%%%%%%%%%%%%%%%%%%%%%%%%%%%%%%%

\subsubsection[]{Radio spectrum}\label{ssec:radiospec}

\begin{figure}
\centering
\includegraphics[scale=0.53]{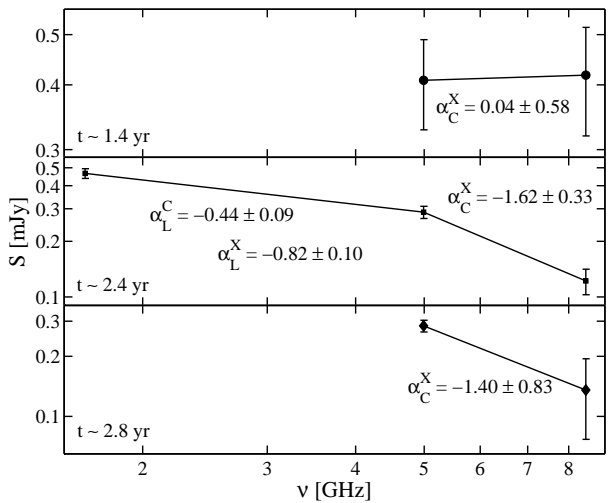}
\caption{Radio continuum spectrum of SN\,2010P at 1.4, 2.4 and 2.8\,yr after
explosion.} \label{fig:radioSED}
\end{figure}

We have obtained a radio continuum spectrum of SN\,2010P at three different ages (Figure \ref{fig:radioSED}). 
We used the corrected flux densities retrieved from the quasi-simultaneous epochs E4 and E5 at 1.4\,yr, E6 to E8
at 2.4\,yr, and from epochs E9 and E10 at an age of 2.8\,yr. 

%%%%%%%%%%%%%%%%%%%%%%%%%%%%%%%%%%%%%%%%%%%%%%%%%%%%%%%%%%%%%%%%%%%%%%%%%%%%%%%%%%
%%%%%%%%%%%%%%%%%%%%%%%%%%%%%%%%%%%%%%%%%%%%%%%%%%%%%%%%%%%%%%%%%%%%%%%%%%%%%%%%%%

At 1.4\,yr, the two-point spectral index $\left( S_{\nu} \propto \nu^{\alpha} \right)$ between 5.0 and 8.5\,GHz is 
$\alphxc(1.4\rmn{yr})=0.04\pm 0.58$, using the corrected values from Table \ref{tab:fluxcorr}. The resulting 1.7 to 
5.0\,GHz and 5.0 to 8.5\,GHz spectral indices at 2.4\,yr are $\alphcl(2.4\rmn{yr})= -0.44 \pm 0.09$, and 
$\alphxc(2.4\rmn{yr})= -1.62 \pm 0.33$, respectively. At an age of 2.8\,yr, the 5.0 to 8.5\,GHz spectral index 
results in $\alphxc(2.8\rmn{yr})=-1.40\pm 0.83$. 

The flat spectral index that we observe at 1.4\,yr indicates that the SN emission is becoming transparent at 
$\sim$500 days after explosion. We note that once a SN reaches its maximum (at time \tpeak{}) and enters its 
optically thin phase, a higher flux density is observed at lower frequencies \citep{weiler02}, as we observe for 
SN\,2010P at 2.4 and 2.8\,yr. The knee in the spectrum at 2.4\,yr would be a consequence of the radio emission being 
transparent at 5.0 and 8.4\,GHz, whilst still being slightly absorbed at 1.7\,GHz.

We note that $\alphxc(2.4\rmn{yr})$ and $\alphxc(2.8\rmn{yr})$ are highly consistent. However, the latter 
involves a measurement affected by the galactic background emission unlike the former, and hence it has a 
larger uncertainty. We thus consider that $\alphxc(2.4\rmn{yr})$ describes the optically thin phase of the 
SN more reliably (see Section \ref{sec:rlc}).

\subsection[]{The radio light curve of SN\,2010P}\label{sec:rlc}

The shape of a SN light curve is dominated by two main factors: the non-thermal radio emission which
declines slowly, and the rapid decline of non-thermal and thermal absorption, mainly,
synchrotron self-absorption (SSA) and free-free absorption (FFA). SSA originates at the shocked-CSM 
region, when the pressure is high enough so that the ultra-relativistic electrons absorb their own 
synchrotron photons. FFA arises from the ionized medium surrounding the SN, which absorbs the synchrotron 
photons.

There are a number of models in the literature which allow a successful parametrisation of SN light curves 
\citep[see e.g.,][]{weiler02, soderberg05}. Whilst these models include many details on the characteristics 
of the SN itself and its CSM, a good sampling of the radio emission through time at multiple frequencies is 
required to constrain all of these parameters. 

We make an attempt to parametrise the light curve of SN\,2010P using the model described in \citet{weiler02}. 
Given that the sampling we have is very limited, we aim to obtain a qualitative view of the SN\,2010P radio 
behaviour when comparing it with other known radio SNe. At the late SN stages covered by our data, we expect 
that the radio light curves will be dominated by FFA, however, as SSA has been found to be important in 
the evolution of type Ib/IIb SNe, here we explore both pure SSA and pure FFA models. Whilst a combined 
SSA and FFA model would be ideal, we note that this is not feasible due to the scarce existing data,
specially at early epochs covering the optically thick part of the SN evolution.

Following \citet{weiler02}, the evolution in time ($t$) of the flux density ($S$) at a frequency ($\nu$) is 
given by,
\begin{equation}\label{eq:paramweiler}
S[\mjy] = K_1\left(\frac{\nu}{5~\rmn{GHz}} \right)^{\alpha}\left(\frac{t-t_0}{1~\rmn{day}}\right)^{\beta}A_1~A_2~A_3
\end{equation}
with 
\[ A_1 = e^{-\tau_{\rmn{external}}};\]
\[ A_2 = \frac{1-e^{-\tau_{\rmn{CSM}_{\rmn{clumps}}}}}{\tau_{\rmn{CSM}_{\rmn{clumps}}}},\]
where 
\[ \tau_{\rmn{external}}=  K_2\left(\frac{\nu}{5~\rmn{GHz}} \right)^{-2.1}\left(\frac{t-t_0}{1~\rmn{day}}\right)^{\delta}  +
                                   K_4\left(\frac{\nu}{5~\rmn{GHz}} \right)^{-2.1}\]
is the optical depth of homogeneous CSM, whereas that corresponding to clumpy CSM is
\[ \tau_{\rmn{CSM}_{\rmn{clumps}}}=K_3\left(\frac{\nu}{5~\rmn{GHz}} \right)^{-2.1}\left(\frac{t-t_0}{1~\rmn{day}}\right)^{\delta^{\prime}},\]
with $\delta^{\prime}=\frac{5}{3}\delta$; and
\[ A_3 = \frac{1-e^{-\tau_{\rmn{internal}_{\rmn{SSA}}}}}{\tau_{\rmn{internal}_{\rmn{SSA}}}},\]
where 
\[ \tau_{\rmn{internal}_{\rmn{SSA}}}=  K_5\left(\frac{\nu}{5~\rmn{GHz}} \right)^{\alpha-2.5}\left(\frac{t-t_0}{1~\rmn{day}}\right)^{\delta^{\prime\prime}},\]
is the optical depth for SSA. 

The time dependence of the supernova emission is described by $\beta$, whereas $\delta$ and 
$\delta^{\prime\prime}$ describe the time dependence of the optical depth for the absorption by the local 
ionized medium and by SSA, respectively. The explosion date is denoted by $t_0$, and the spectral index in 
the optically thin phase of the SN by $\alpha$. The $K$-terms correspond to the flux density ($K_1$), the 
absorption by a homogeneous ($K_2$ and $K_4$) and/or clumpy ($K_3$) CSM, and the internal non-thermal 
absorption ($K_5$) at a frequency of 5\,GHz. 

In the following, we adopt $t_0=$ 2010 January 10 (from \citeauthor{kankare13}), $\alpha= -1.62\pm0.33$ 
(from Section \ref{ssec:radiospec}).

\subsubsection[]{Pure SSA model and results}\label{ssec:SSA}

In the pure SSA model the absorption by the ionized CSM is negligible, so $A_1$ and $A_2 \to 1$ in
Equation \ref{eq:paramweiler},
and we only need to solve for $K_1$, $K_5$, $\beta$ and $\delta^{\prime\prime}$. We do this by implementing 
a Monte Carlo simulation as explained in Appendix \ref{ap:mcexpt}. This resulted in $K_1 = 2.64_{-0.19}^{+0.17}\e3$, 
$K_5 = 8.20_{-0.82}^{+0.80}\ee12$, $\beta = -1.34_{-0.01}^{+0.01}$ and 
$\delta^{\prime\prime} = -4.80_{-0.01}^{+0.01}$, with a reduced $\chi^2$ of 5.2. The fitted values were 
used to draw the parametrised light curves we present in Figure \ref{fig:lc_SSA}, which provides
approximate estimates for the peak luminosities at each frequency and the time at which these occurred 
(see Table \ref{tab:SSA_param}). We note that the inferred peak flux densities at lower frequencies seem 
to be much higher than those at higher frequencies. This might be the effect of the absorption decreasing 
very rapidly (as indicated by the high, negative value of $\delta^{\prime\prime}$) and thus affecting less 
the lower frequencies than the higher ones.

To infer physical parameters characterising the SN and its progenitor, we will consider in the following
that the power-law distribution of relativistic electron energies is governed by $p=3$ ($N(E) \propto 
E^{-p}$) and assume both that the ratio of the magnetic energy density to that of particles is equal to
one (i.e., equipartition), and that the emitting region fills half of the volume determined by a spherical 
blast wave. We will concentrate on the physical parameters obtained at 5.0 and 8.5\,GHz since the radio 
evolution of SN\,2010P is better constrained at these frequencies, compared to 29.0 and 1.7\,GHz, where 
only one data point is available.

When SSA is the dominant absorption mechanism, the radius of the blast wave according to \citet{chevalier98}
can be obtained from 
\[ R_{\rmn{shell}} = 8.8\ee15 \left(\frac{\speak{}}{\rmn{Jy}}\right)^{\frac{9}{19}} 
\left(\frac{D}{\rmn{Mpc}}\right)^{\frac{18}{19}} \left( \frac{\nu_{\rmn{peak}}}{5~ \rmn{GHz}}\right)^{-1} \rmn{cm},\]
where $D$ is the distance to the SN (44.8\,Mpc). Plugging in the values from Table \ref{tab:SSA_param}, we 
obtain $R_{\rmn{shell}} = (8.3, 4.3)\ee15$\,cm for $\nu_{\rmn{peak}} =$ 5.0 and 8.5\,GHz,
respectively. These values are well below the upper limit set by our EVN observations in epoch E6 ($\sim$4.3\ee17\,cm; 
Section \ref{ssec:radem}). For the different frequencies, we then infer a brightness temperature of $\sim$3.5\ee10\,K
at both frequencies,
which is consistent with non-thermal emission, close to, but without surpassing the Inverse Compton Catastrophe 
limit \citep[$\tb^{\rmn{ICC}} \sim 1\ee11$\,K, e.g.,][]{readhead94} above which equipartition no longer holds. 
We also calculate average expansion speeds from $R_{\rmn{shell}}/\tpeak{}$, leading to sub-relativistic velocities 
for the radio ejecta of 1,460--1,785\,\kms{} (or $\sim$0.005$c$).

We can estimate the deceleration parameter ($m$) considering that the blast-wave radius evolves in time as 
$R_{\rmn{shell}} \propto t^m$. A basic line fitting in the $R_{\rmn{shell}}$ vs. \tpeak{} plot yields $m=1.4$. 
This is well outside the possible range of values that guarantee the existence of a self-similar solution for 
the post-shock flow \citep[$2/3\la m\la 1$, e.g.,][]{chevalier82}, with $m<1$ meaning that the shock is decelerating, 
and $m=1$ remaining steady. Another possibility is to use $m=-(\alpha - \beta -3)/3$ \citep{weiler02}, which 
results in $1.10\pm0.11$ and represents a permitted value within the uncertainties, but which might also 
indicate that at least one of the assumptions we have made does not apply.

The magnetic field strength at the time of the peak under the assumptions above, can be calculated as
\[ B_{\rmn{peak}} = 0.58 \left(\frac{\speak{}}{\rmn{Jy}}\right)^{\frac{-2}{19}} 
\left(\frac{D}{\rmn{Mpc}}\right)^{\frac{-4}{19}} \left( \frac{\nu_{\rmn{peak}}}{5~ \rmn{GHz}}\right) \rmn{G}\]
following \citet{chevalier98}, and resulting in 0.6--1.0\,G for SN\,2010P.

In addition to the assumptions made above, we also consider a minimum Lorentz factor $\gamma_{\rmn{m}} =1$, so that 
the electron density \citep{horesh13},
\[ n_e = \frac{1}{2} \left(\frac{B_{\rmn{peak}}^2}{8\pi m_e c^2}\right)\,\rmn{cm}^{-3}\]
is (0.8, 2.5)\e4\,cm$^{-3}$ at 5.0 and 8.5\,GHz. This allows us to
obtain an estimate of the ratio of the mass-loss rate in units of \ml{} ($\dot M$) to wind 
velocity of the progenitor star in units of 10\,\kms{} ($v_{\rmn{wind10}} = v_{\rmn{wind}}/10\,\kms{}$),
\[ \frac{\dot M}{v_{\rmn{wind10}}}  =  1.6\eee-20 \times 4 \pi \left( \frac{R_{\rmn{shell}}}{\rmn{cm}}\right)^2 
   \left( \frac{m_p n_e}{\rmn{g}~ \rmn{cm}^{-3}}\right), \]
which for SN\,2010P yields (1.6--1.9)\ee-7.

%%%%%%%%%%%%%%%%%%%%%%%%%%%%%%%%%%%%%%%%%%%%%%%%%%%%%%%%%%%%%%%%%%%%%%%%%%%%%%%%%%
\begin{center}
\begin{table}
\centering
\caption{\protect{ Approximate peak values from the pure SSA fit. }} \label{tab:SSA_param}
\begin{tabular} {cccc} \hline
\multicolumn{1}{c}{Frequency} & \multicolumn{1}{c}{\tpeak{}} & \multicolumn{1}{c}{\speak{}}  &  \multicolumn{1}{c}{\lpeak{}}\\
\multicolumn{1}{c}{(GHz)} & \multicolumn{1}{c}{(days)} & \multicolumn{1}{c}{(\mujy{})} & \multicolumn{1}{c}{(\lunits{})} \\
    \hline
1.7  & 1392 & 743 & 1.8\ee27 \\
5.0  & 538  & 440 & 1.1\ee27 \\
8.5  & 344  & 342 & 8.2\ee26 \\
29.0 & 120  & 191 & 4.6\ee26 \\
    \hline 
   \end{tabular}
\end{table}
\end{center}
%%%%%%%%%%%%%%%%%%%%%%%%%%%%%%%%%%%%%%%%%%%%%%%%%%%%%%%%%%%%%%%%%%%%%%%%%%%%%%%%%%

\subsubsection[]{Pure FFA model and results}\label{ssec:FFA}

Just as for the SSA model, we performed a Monte Carlo simulation to solve for $K_1$, 
$K_2$, $\beta$ and $\delta$, having adopted the terms $K_3$ and $K_4$ to be sufficiently small so 
that \[ \tau_{\rmn{external}} =  K_2\left(\frac{\nu}{5~\rmn{GHz}} \right)^{-2.1}
\left(\frac{t-t_0}{1~\rmn{day}}\right)^{\delta}\]
and $A_2 \to 1$ in Equation \ref {eq:paramweiler}(see the discussion on these approximations and 
details on how the fit was made in Appendix \ref{ap:mcexpt}). In the pure FFA model, SSA is 
not relevant, so $A_3 \to 1$ as well. 

\begin{figure*}
\centering
\subfigure[]{
 \includegraphics[width=\columnwidth]{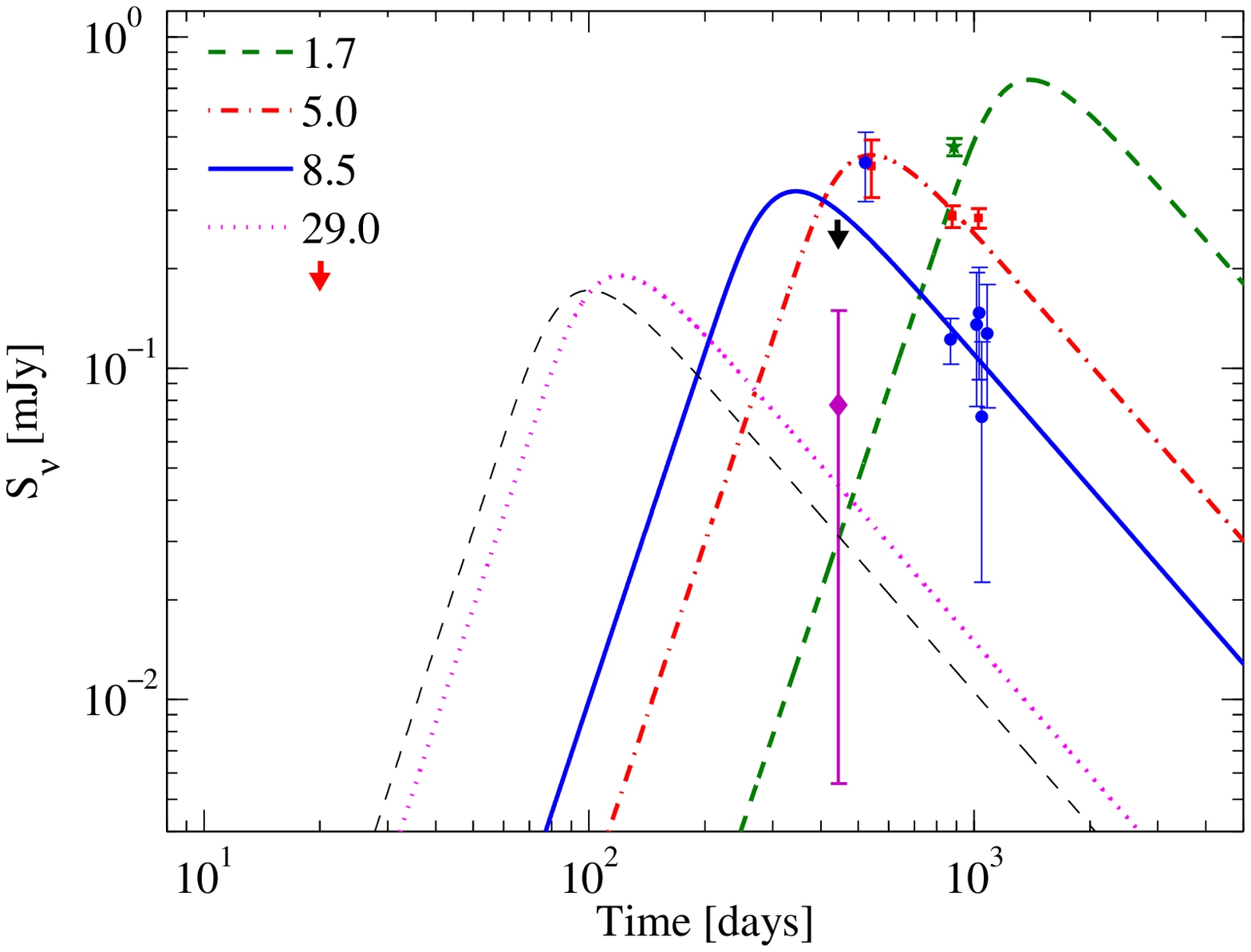} \label{fig:lc_SSA}}
 \subfigure[]{
 \includegraphics[width=\columnwidth]{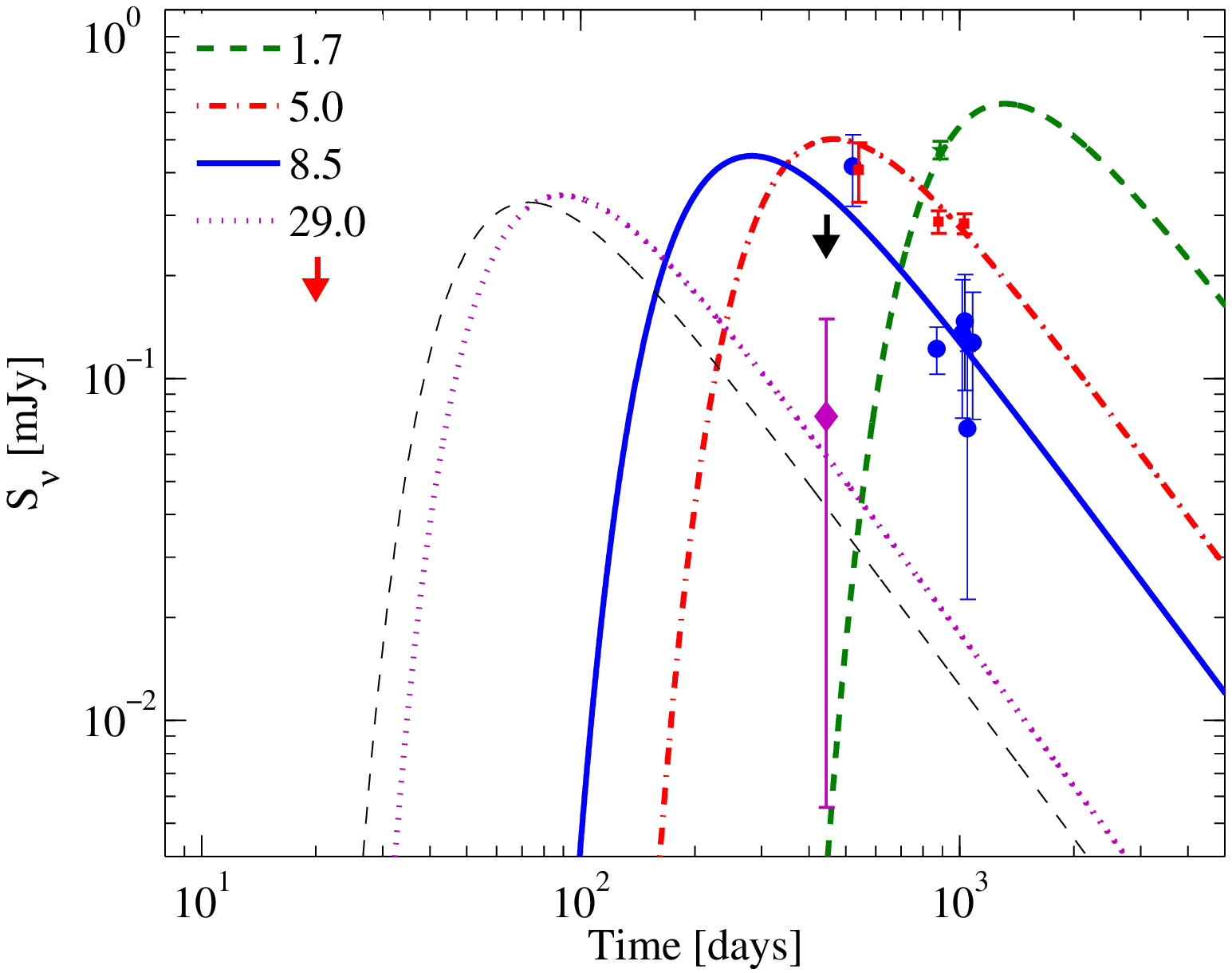} \label{fig:sn2010pLC}}
 \caption{Parametrised radio light curve of SN\,2010P {\bf --\subref{fig:lc_SSA} pure SSA and 
         \subref{fig:sn2010pLC} pure FFA--} at 1.7\,GHz (green star, dashed line), 5.0\,GHz 
         (red squares, dash-dotted line), 8.5\,GHz (blue circles, solid line) and 29\,GHz 
         (magenta diamond, dotted line). The downward arrows represent 3$\sigma$ upper limits 
         from epochs E1 (red), and E3 (black). The black dashed line is the predicted light curve 
         at 36.0\,GHz. }          
\end{figure*}

The Monte Carlo simulation resulted in $K_1 = 9.20_{-1.15}^{+4.12}\e3$, $K_2 = 5.68_{-0.72}^{+0.98}\e5$, 
$\beta = -1.49_{-0.05}^{+0.01}$ and $\delta = -2.22_{-0.02}^{+0.02}$, with a reduced $\chi^2$ of 2.6.
The fitted light curve (shown in Figure \ref{fig:sn2010pLC}) reproduces the spectral behaviour we 
infer from our spectral index measurements at 2.4 and 2.8\,yr (see Figure \ref{fig:radioSED}). In Table 
\ref{tab:FFA_param} we show the approximate time to peak, peak flux density and peak luminosities 
at the different frequencies, obtained from the FFA light-curve fit (Figure \ref{fig:sn2010pLC}).

%%%%%%%%%%%%%%%%%%%%%%%%%%%%%%%%%%%%%%%%%%%%%%%%%%%%%%%%%%%%%%%%%%%%%%%%%%%%%%%%%%

%%%%%%%%%%%%%%%%%%%%%%%%%%%%%%%%%%%%%%%%%%%%%%%%%%%%%%%%%%%%%%%%%%%%%%%%%%%%%

Using the time dependence of the absorption due to the ionized CSM, the deceleration parameter is 
calculated as $m=-\delta/3$, resulting in $0.74 \pm 0.01$ for SN\,2010P. In the FFA case at peak times, 
if assuming that the ejecta velocity can be measured from optical lines (e.g., H$\alpha$) 45\,days after 
explosion \citep{weiler86}, we have that
\[ R_{\rmn{shell}} = \frac{3.89\ee15}{m} \left( \frac{v_{\rmn{shock}}}{1\e4\,\kms{}}\right) \left( 
  \frac{\tpeak}{45\,\rmn{days}}\right)^m \rmn{cm}.\]

SN ejecta is expected to have a distribution in velocity as it moves through the CSM. For example, for the
Type IIb SN\,2011hs, $v_{\rmn{shock}}$ ranged from 8,000 to 12,000\,\kms{} \citep[][]{bufano14}. Recently,
\citet{inserra13} investigated the photometric properties of a group of moderately luminous Type II SNe
and found that in the early phases of the SN explosion, the expansion velocities as measured from H$\alpha$ 
and H$\beta$, range between $10,000$--$12,000$\,\kms{}. For SN\,2010P we use the commonly used shock velocity 
value for Type II SNe $v_{\rmn{shock}}\approx 10,000$\,\kms{} near maximum light \citep[e.g.,][]{weiler02}, 
as we lack measured shock velocity values (\citeauthor{kankare13}). Using the $m$ value we estimated above 
and the \tpeak{} values from Table \ref{tab:FFA_param}, we obtain $R_{\rmn{shell}} = (29.6, 20.3)\ee15$\,cm 
for $\nu_{\rmn{peak}} =$ 5.0 and 8.5\,GHz, respectively. These sizes do not violate the EVN limit of 4.3\ee17\,cm, 
and imply brightness temperatures of (3.2, 2.1)\e9\,K, and thus below the ICC limit. As we did in the SSA model, 
the average expansion speeds in this case are 7,380--8,410\,\kms{} (or 0.02--0.03$c$), hence, also sub-relativistic. 

To estimate the progenitor's average mass-loss rate, we can follow two approaches. For the first approach,
we consider equation (11) in \citet{weiler02} and we assume that only the uniform component of the CSM 
contributes to the absorption of SN\,2010P radio emission, so that at day \tpeak{}, the ratio of $\dot M$ 
to $v_{\rmn{wind10}}$ is given by

\begin{eqnarray*}
\frac{\dot M}{v_{\rmn{wind10}}} &=& 3\ee-6 ~\langle \tau_{\rmn{eff}}^{0.5}\rangle ~ m^{-1.5}~
  \left(\frac{v_{\rmn{shock}}}{10^4\,\kms}\right)^{1.5}  \\
 & \times & \left( \frac{\tpeak}{45\,\rmn{days}}\right)^{1.5m} \left(\frac{T}{10^4\,\rmn{K}}\right)^{0.68}
\end{eqnarray*}
\noindent with $\langle \tau_{\rmn{eff}}^{0.5}\rangle = \tau_{\rmn{external}}^{0.5}$, which for SN\,2010P is 
$\sim 0.82$ at all the considered frequencies. Assuming that $v_{\rmn{shock}}\approx10,000$\,\kms{}, and 
10,000\,K for the CSM temperature $T$ (corresponding to the equilibrium temperature of a typical \hii{} region), 
we obtain $\dot M/v_{\rmn{wind10}}=(5.1, 3.0)\ee-5$ at 5.0 and 8.5\,GHz, respectively.

Alternatively, a second approach would be to follow \citeauthor{weiler02}'s equation (18) for Type II SNe 
at 5\,GHz:
\begin{eqnarray*}
\frac{\dot M}{v_{\rmn{wind10}}}  =  1\ee-6 \left(\frac{\lpeak}{10^{26}\,\lunits{}} \right)^{0.54} 
    \left(\frac{\tpeak}{\rmn{days}} \right)^{0.38}
\end{eqnarray*}
we obtain $\dot M/v_{\rmn{wind10}} < 4.0\ee-5$, consistent with the upper limit obtained following the first 
approach. We note that the relation used as a second approach does not depend on $v_{\rmn{shock}}$ and thus,
assuming that $\dot M/v_{\rmn{wind10}}$ is equally good following the first and the second approach, 
we can plug in the ratio obtained here at 5\,GHz into the first relation to solve for $v_{\rmn{shock}}$. We 
obtain $v_{\rmn{shock}}\sim 8,400$\,\kms{} at 5\,GHz, which matches quite well with the average expansion speed 
we estimated above. 
From now on, we consider $\dot M/v_{\rmn{wind10}}=(3.0$--5.1)$\ee-5$ as a representative value for the 
mass-loss rate to wind velocity ratio.

We can now also estimate the CSM density as
\begin{eqnarray*}
\rho_{\rmn{CSM}} &=& 1.6\e6 \left( \frac{\dot M}{1\ee-5 \ml{}} \right) \left( \frac{v_{\rmn{wind}}}{10\,\kms{}}\right)^{-1}\\
& \times &  \left(\frac{R_{\rmn{shell}}}{1\ee15 \rmn{cm}}\right)^{-2} ~ \rmn{g ~cm}^{-3}
\end{eqnarray*} 
\citep[e.g.,][]{horesh13}. For SN\,2010P, this leads to 
$\rho_{\rmn{CSM}}=$ (0.7, 1.2)\e4\,g~cm$^{-3}$ at 5.0 and 8.5\,GHz, respectively.

%%%%%%%%%%%%%%%%%%%%%%%%%%%%%%%%%%%%%%%%%%%%%%%%%%%%%%%%%%%%%%%%%%%%%%%%%%%%%%%%%%
\begin{center}
\begin{table}
\centering 
\caption{\protect{ Approximate peak values from the pure FFA fit. }} \label{tab:FFA_param}
\begin{tabular} {cccc} \hline
\multicolumn{1}{c}{Frequency} & \multicolumn{1}{c}{\tpeak{}} & \multicolumn{1}{c}{\speak{}}  &  \multicolumn{1}{c}{\lpeak{}}\\
\multicolumn{1}{c}{(GHz)} & \multicolumn{1}{c}{(days)} & \multicolumn{1}{c}{(\mujy{})} & \multicolumn{1}{c}{(\lunits{})} \\
    \hline
1.7  & 1307 & 637 & 1.5\ee27 \\
5.0  & 464  & 502 & 1.2\ee27 \\
8.5  & 280  & 448 & 1.1\ee27 \\
29.0 &  88  & 343 & 8.2\ee26 \\
    \hline 
   \end{tabular} 
\end{table}
\end{center}
%%%%%%%%%%%%%%%%%%%%%%%%%%%%%%%%%%%%%%%%%%%%%%%%%%%%%%%%%%%%%%%%%%%%%%%%%%%%%%%%%%

\subsubsection[]{FFA vs. SSA}\label{ssec:FFAvsSSA}

In Sections \ref{ssec:SSA} and \ref{ssec:FFA} we have obtained physical parameters 
for SN\,2010P assuming pure SSA and pure FFA models dominating its radio emission. 

The  reduced $\chi^2$ of the SSA model doubles that of the FFA model, 
indicating that FFA describes better the late-time observations we present here. Beyond 
this, we also note some inconsistencies in the inferred parameters from the SSA model. 
For instance, following two different approaches we found that the deceleration 
parameter is $\ga 1$ in the pure SSA model. This clashes with the expectation of rapid 
deceleration as the SN shock moves into a dense CSM, and implies that the blast wave has 
been expanding at average shock velocities $<2,700\,\kms$ (Section \ref{ssec:SSA}). These 
velocities are unrealistically low, whereas those obtained for FFA are in fact physically 
reliable and agree better with measured velocities for other SNe \citep{inserra13}
(see Section \ref{ssec:FFA}). Furthermore, the low mass-loss to wind velocity ratio
we obtained in the pure SSA model ($\dot M [\ml{}] / v_{\rmn{wind}} [10\,\kms{}]=$(1.6--1.9)$\ee-7$), 
is difficult to reconcile with the high luminosity of SN\,2010P. 

Whilst SSA might be important at early stages of SN\,2010P (where the existing data is scarce), 
we find that the FFA model fits the available data better and we thus adopt FFA as the dominant 
absorption mechanism for SN\,2010P.

\subsubsection[]{SN\,2010P characterisation and radio parameters}

In Table \ref{tab:SNcomp} we show the fitting parameters for SN\,2010P we obtained when
assuming FFA (Section \ref{ssec:FFA}), together with those of other well-studied Type II radio 
SNe from the literature for comparison purposes. We only included those fitting parameters relevant 
for comparison with SN\,2010P.

From Table \ref{tab:SNcomp} we note that Type IIb SNe have the steepest spectral indices among the 
Type II SNe, as well as the lowest mass-loss rates, and this is the case as well for SN\,2010P. Other 
parameters such as $K_1$ and $\beta$ vary greatly among SNe of the same type, thus cannot be used to 
set further constraints. The value for parameter $\delta$ tends to be smaller for Type IIL and IIn SNe, 
and the one we obtained for SN\,2010P is typical for Type IIb SNe, with the exception of SN\,2011hs, whose 
$\delta$ is significantly different from the value fitted for other Type IIb SNe, however, this is consistent 
with the rapid deceleration observed in the time evolution of the ejecta velocity (\citealt{bufano14}).

Parameter $K_2$ is indeed much larger for SN\,2010P compared to other Type IIb SNe. However, we note 
that such a high value has also been obtained for SN\,2001gd when assuming a pure FFA model for its radio 
light curve \citep{stockdale03}, as we have assumed for SN\,2010P. \citet{stockdale07} improved on their 
results from \citeyear{stockdale03} by gathering a more complete data set of observations, and combining 
SSA and FFA components in the parametrization of the SN\,2001gd light curves. 

%%%%%%%%%%%%%%%%%%%%%%%%%%%%%%%%%%%%%%%%%%%%%%%%%%%%%%%%%%%%%%%%%%%%%%%%%%%%%
\begin{table*}
\centering 
\caption{\protect{Fitting parameters from radio light-curve models of SN\,2010P and known Type II SNe.}} 
\label{tab:SNcomp} 
\begin{tabular}{ccccccccc} \hline
SN & $\alpha$ & $\beta$ & $\delta$ & $K_1$  & $K_2$ &  $\tpeak$ & $\lpeak$    & $\dot M/v_{\rmn{wind10}}$   \\
~  &   ~      &   ~     &    ~     & (mJy) &  ~   &   (days)  & (\lunits{}) &  \\
\hline				
2010P (IIb)$^a$  & $-1.62\pm0.33$ & $-1.49_{-0.05}^{+0.01}$ & $-2.22_{-0.02}^{+0.02}$ & $9.20_{-1.15}^{+4.12}\e3$ & $5.68_{-0.72}^{+0.98}\e5$ &  $\sim464$ & $\sim1.2\ee27$ & ($3.0$--$5.1)\ee-5$   \\		     
1993J (IIb)$^b$  & $-0.81$ & $-0.73$ & $-1.88$ & $4.8\e3$  & $1.6\e2$  & 133 & $1.5\ee27$ & (0.5-5.9)$\ee-5$ \\
2001gd (IIb)$^c$ & $-0.94$ & $-0.92$ & $-1.88$ & $7.50\e2$ & $1.50\e3$ &  80 & $3.8\ee27$ & $(0.7-5.6)\ee-5$    \\
2001ig (IIb)$^d$ & $-1.06$ & $-1.50$ & $-2.56$ & $2.71\e4$ & $1.38\e3$ &  74 & $3.5\ee27$ & $(2.2\pm0.5)\ee-5$    \\
2011hs (IIb)$^e$ & $-1.90$ & $-1.66$ & $-1.31$ & $5.2\e3$  & $1.5\e2$  &  59 & $1.6\ee27$ & $(2.0\pm0.6)\ee-5$  \\
1970G (IIL)$^f$  & $-0.55$ & $-1.87$ & $-3.00$ & $1.77\e6$ & $1.80\e7$ & 307 & $1.4\ee27$ & $6.8\ee-5$    \\
1979C (IIL)$^f$  & $-0.75$ & $-0.80$ & $-2.94$ & $1.72\e3$ & $3.38\e7$ & 556 & $2.6\ee27$ & $1.1\ee-4$    \\
1980K (IIL)$^f$  & $-0.60$ & $-0.73$ & $-2.69$ & $1.15\e2$ & $1.42\e5$ & 134 & $1.2\ee26$ & $1.3\ee-5$    \\
1978K (IIn)$^g$  & $-0.77$ & $-1.55$ & $-2.22$ & (0.2-76)$\e7$&(0.3-90)$\e4$&940-1000&(0.9-2.0)$\ee28$&$1.0\ee-4$ \\
1986J (IIn)$^f$  & $-0.66$ & $-1.65$ & -       & $1.19\e7$ & -         & 1210 & $2.0\ee28$ & $4.3\ee-5$    \\
1988Z (IIn)$^f$  & $-0.69$ & $-1.25$ & -       & $1.47\e4$ & -         &  898 & $2.3\ee28$ & $1.1\ee-4$    \\
2000ft (IIn)$^h$ & $-1.27$ & $-2.02$ & $-2.94$ & $4.45\e5$ & $1.67\e7$ &  300 & $1.1\ee28$ & (4.7-5.1)\ee-5\\
    \hline
   \end{tabular} 
\begin{flushleft}
\vspace{-1mm}
This is not a complete list of fitting parameters. 
References: (a) This work, (b) \citealt{weiler07}, (c) \citealt{stockdale07}, (d) \citealt{ryder04}, (e) \citealt{bufano14}, 
(f) \citealt{weiler02}, (g) \citealt{schlegel99}, (h) \citealt{perez09a}. 
\end{flushleft}
\end{table*}
%%%%%%%%%%%%%%%%%%%%%%%%%%%%%%%%%%%%%%%%%%%%%%%%%%%%%%%%%%%%%%%%%%%%%%%%%%%%%%%%%%

\begin{figure}
\centering
\includegraphics[width=84mm]{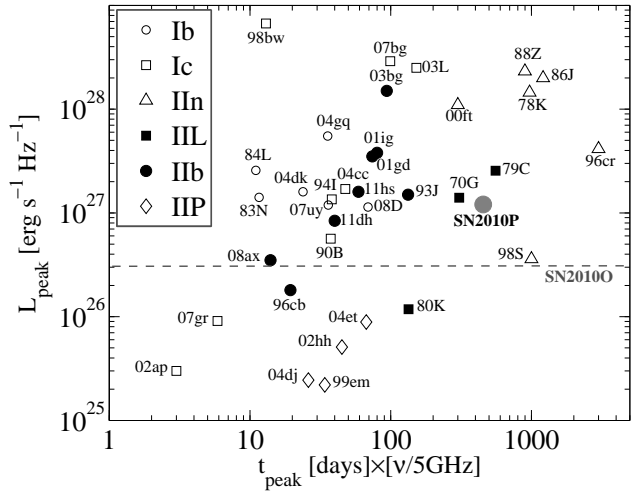}
\caption{Peak monochromatic luminosity vs. time to peak from explosion date for different types of CCSNe
(hypernovae and other peculiar objects included). The first peak is used in case more than one has been 
reported. The dashed line represents the upper limit in luminosity for SN\,2010O. References: 
\citet{weiler02} for SNe 1970G, 1979C, 1980K, 1983N, 1984L, 1986J, 1988Z, 1990B and 1994I; 
\citet{pooley02} for 1998S and 1999em; \citet{wellons12} for 2004cc, 2004dk and 2004gq; 
\citet{horst11} for 2007uy and 2008D; \citet{schlegel99} for 1978K;
\citet{weiler07} for 1993J; \citet{weiler98} for 1996cb; \citet{bauer08}  for 1996cr; 
\citet{weiler01} for 1998bw; \citet{perez09a} for 2000ft; \citet{stockdale07} for 2001gd; 
\citet{ryder04} for 2001ig; \citet{berger02} for 2002ap; \citet{chevalier06} for 2002hh;  
\citet{soderberg05} for 2003L; \citet{soderberg06} for 2003bg; \citet{beswick05} for 2004dj; 
\citet{marti07} for 2004et; \citet{salas13} for 2007bg; \citet{soderberg10} for 2007gr; 
\citet{roming09} for 2008ax; \citet{maeda12} for 2011dh \citep[see also][]{horesh13}; \citet{bufano14} 
for 2011hs.}
\label{fig:LumTime}
\end{figure}

The fitted light curves we present in Figure \ref{fig:sn2010pLC} assuming FFA, reproduce the optically thin 
behaviour we infer from our spectral index measurements at 2.4 and 2.8\,yr (see Figure \ref{fig:radioSED}), and 
provide a rough estimate for the time it took the SN to reach its peak at 5\,GHz ($\tpeak\sim464$\,days) and the
corresponding peak luminosity at 5\,GHz ($\lpeak\sim1.2\ee27$\lunits{}), which can be used to infer the 
SN type (see Figure \ref{fig:LumTime}). We note that SN\,2010P lies close to Type IIL SNe in that plot. However, 
the early optical spectrum of SN\,2010P (\citeauthor{kankare13}) does not show strong hydrogen features, as should 
be observed in a typical Type IIL SN.

%%%%%%%%%%%%%%%%%%%%%%%%%%%%%%%%%%%%%%%%%%%%%%%%%%%%%%%%%%%%%%%%%%%%%%%%%%%%%%%%%%
%%%%%%%%%%%%%%%%%%%%%%%%%%%%%%%%%%%%%%%%%%%%%%%%%%%%%%%%%%%%%%%%%%%%%%%%%%%%%%%%%%

\section{Discussion}\label{sec:discussion}

The optical spectroscopy reported in \citeauthor{kankare13} indicates that 2010O is a Type Ib SN, 
which are known to be bright at radio frequencies. In fact a peak luminosity $<3\ee26$\,\lunits{} 
(see Figure \ref{fig:LumTime}) is quite uncommon for Type Ib/c SNe \citep{soderberg07}. Additionally, 
these SNe evolve very rapidly and thus reach their peak luminosity within a few tens of days \citep[e.g., 
SN\,2008D,][]{soderberg08}, so the most likely explanation for the radio non-detection of SN\,2010O is that 
it reached its peak some time after the early-time MERLIN observations by \citeauthor{beswick10}, and well 
before our observations in 2011, so it went unnoticed owing to the lack of prompt radio observations. 

SN\,2010P lies in a location with larger extinction than seen toward SN\,2010O ($A_V =7$\,mag, 
\citeauthor{kankare13}) and at a $\approx 160$\,pc projected distance from the component C$^{\prime}$, 
SN\,2010P is the first radio SN detected in the outskirts of a bright radio emitting component in 
Arp\,299, despite the intense radio monitoring. Its strong radio emission 1.4 to 2.8\,yr after explosion 
indicates a strong interaction between the SN ejecta and the CSM. Whilst the early optical spectrum is 
consistent with the SN being either a Type Ib or IIb (\citeauthor{kankare13}), the late-time radio detection 
rules out a Type Ib origin.

The parameters we obtained to fit the radio light curve of SN\,2010P are common for luminous Type II SNe (see 
Table \ref{tab:SNcomp}). Whilst the inferred peak luminosity and the time to reach the peak are more comparable 
to those of Type IIL's, its estimated mass-loss rate to wind velocity ratio, 
$\dot M/v_{\rmn{wind10}}=(3.0$--$5.1)\ee-5$, and its inferred deceleration parameter ($\sim0.74$), 
support better a Type IIb nature with a slow evolution, and are consistent with the early optical spectra lacking 
strong hydrogen features (\citeauthor{kankare13}). We note that some Type IIn SNe have comparable mass-loss rate to 
wind velocity ratios (see Table \ref{tab:SNcomp}), however, these SNe are also an order of magnitude more luminous 
than SN\,2010P at their peak.

It has been suggested that Type IIb SNe can be divided into two broad groups \citep{chevalier10} under the 
assumption that SSA is the main absorption mechanism: i) slowly evolving SNe at radio frequencies ($\tpeak>100$\,d) 
with a large ($>0.1\,\msun{}$) hydrogen mass envelope and slow ejecta velocity ($\sim 10,000\,\kms{}$), expected from 
an extended progenitor ($R\sim10^{13}$\,cm); ii) rapidly evolving SNe ($\tpeak<100$\,d) with a small hydrogen mass 
envelope ($<0.1\,\msun{}$) and fast ejecta velocities (a factor of 3-5 larger than in case i), expected from a compact 
progenitor ($R\sim10^{11}$\,cm). According to this interpretation, SN\,2010P would come from an extended progenitor since 
it took hundreds of days to reach its peak (in both FFA and SSA fits). In \citeauthor{kankare13} we do not find evidence 
for a large hydrogen mass envelope (i.e., $>0.1\,\msun{}$) around SN\,2010P, in contradiction with the expectations for an 
extended progenitor as inferred from its radio behaviour. The properties of SNe 2011dh and 2011hs where more direct evidence 
of the progenitor star has been obtained \citep[][respectively]{bersten13, bufano14}, further suggest that the inferred radio 
properties of a SN are not a strong beacon of the progenitor size. This could be related to the assumption of SSA dominance 
for poorly sampled light curves.

In the case of SN\,2001gd, the assumptions made to fit its radio light curve greatly affected the estimate for the 
time it took the SN to reach its peak at 5\,GHz, going from 173\,d in the FFA model \citep{stockdale03}, to 80\,d 
in the SSA$+$FFA model \citep{stockdale07}, thus placing this SN in the compact IIb category, following the 
interpretation from \citet{chevalier10}. Unlike in the SN\,2001gd case, we do not have early data for SN\,2010P 
describing its turn-on phase at any frequency, and thus, we cannot quantify the presence of SSA in its evolution. 
However, it is rather unlikely that SSA will dominate the absorption for SN\,2010P radio emission when the SN has 
been detected at such late ages. Thus, although the $\tpeak{}$ we estimated for SN\,2010P represents an upper
limit owing to our assumption of FFA being the only absorption mechanism, we have seen that assuming an SSA 
model pushes the peak at different frequencies to even later times and the same interpretation holds.

However, there is another possible scenario which we cannot exclude. SN\,2010P could have reached a first peak 
at early times not sampled by our data, and we could in fact be witnessing a re-brightening of the SN at later times.
In this situation, \tpeak{} would have occurred earlier and SN\,2010P would appear closer to 
other Type IIb SNe in the \lpeak{} vs. \tpeak{} plot.

Observationally, strong variations in the optically thin phase of different types of CCSNe are rather common 
\citep[see table 4 in][]{soderberg06} owing to the complexity of their CSM. For Type IIb SNe, it has been shown 
that variations in the mass-loss of the progenitor (e.g., luminous blue variable-like stars), can create inhomogeneities 
in the CSM and thus modulations in the SN radio light curve \citep[e.g.,][]{kotak06,moriya13}. 

The sparse sampling of the SN\,2010P light curve does not allow us to conclusively determine whether our late-time 
observations match with the forward shock encountering a first, or a second high-density region. Multiple high-density 
regions can be produced by mass-loss episodes, meaning variations in the mass-loss rate throughout the lifetime
of the progenitor star. However, the early optical/NIR data (\citeauthor{kankare13}) do not show evidence for 
interaction with the CSM at the early stages of the SN, i.e., the interaction with the CSM corresponding 
to the end of the progenitor star's lifetime. Therefore, although we cannot rule out completely the possibility of 
the peak radio luminosity corresponding to a secondary mass-loss episode, the observations (both radio and NIR)
strongly suggest that we are witnessing a slow radio evolution of SN\,2010P.

\section{Conclusions}\label{sec:concl}

We report radio observations towards SNe 2010O and 2010P in the LIRG Arp\,299. SN\,2010O was not detected 
throughout our radio monitoring of the host galaxy which lacked observations of the SN at ages between 
0.6\,month and 1.4\,yr, time enough for a Type Ib SN to have have risen to a peak, then decayed. SN\,2010P 
was detected at various frequencies in its transition to, and in its optically thin phase from $\sim$1 to 
$\sim$3\,yr after explosion. Our observations favour FFA as the dominant absorption mechanism controlling 
the radio emission of this SN. We characterise it as a luminous, slowly-evolving Type IIb 
SN with $\dot M [\ml{}] / v_{\rmn{wind}} [10\,\kms{}]=(3.0$--5.1)$\ee-5$. We have also improved on the 
coordinates previously reported for SN\,2010P with a position accuracy better than 1 mas at $\alpha(J2000) 
= 11\h28\m31\fs3605$, $\delta(J2000) = 58\degr33\arcmin49\farcs315$.

SN\,2010P is one of a select group of 12 Type IIb SNe detected at radio wavelengths: for nine of them a 
radio light curve has been obtained (see references in Figure \ref{fig:LumTime}), and three of them have 
only reported radio detections \citep[SNe 2008bo, PTF\,12os and 2013ak reported in][respectively]{stockdale08,
stockdale12, kamble14}.

SN\,2010P is also the most distant Type IIb SN detected so far, and the one that according to our data, 
has taken the longest time to reach its peak. However, a comprehensive, multi-wavelength study covering 
both rise and decline of the SN emission is needed to investigate possible progenitor star scenarios. We note
that the early epochs after shock breakout are crucial, since this when when we can gather more information 
about the progenitors and the dominant absorption mechanism (SSA/FFA) shaping the radio light curves, thus 
helping us to better understand the CCSN phenomenon, and thereby the interplay between massive stars and their 
CSM.

\section*{Acknowledgements}
The authors are grateful to the anonymous referee for a detailed and constructive report which helped to
improve our manuscript. We also thank Edo Ibar for very helpful discussions and suggestions, and Milena 
Bufano for interesting discussions. The research leading to these results has received funding from the 
European Commission Seventh Framework Programme (FP/2007-2013) under grant agreement No. 283393 (RadioNet3).
We acknowledge support from the Academy of Finland (Project 8120503; CRC and SM), Basal-CATA (PFB-06/2007; 
CRC and FEB), CONICYT-Chile under grants FONDECYT 1101024 (FEB) and ALMA-CONICYT FUND Project 31100004 (CRC), 
Iniciativa Cient\'{\i}fica Milenio grant P10-064-F (Millennium Center for Supernova Science) with input from 
``Fondo de Innovaci\'on para la Competitividad del Ministerio de Econom\'{\i}a, Fomento y Turismo de Chile'' 
(FEB), Spanish MINECO Projects AYA2009-13036-CO2-01 and AYA2012-38491-C02-02, co-funded with FEDER funds (RHI, 
MAPT and AA) and the Jenny and Antti Wihuri Foundation (EK).

%%%%%%%%%%%%%%%%%%%%%%%%%%%%%%%%%%%%%%%%%%%%%%%%%%%%%%%%%%%%%% REFERENCES

\appendix

\section{Radio light-curve fit}\label{ap:mcexpt}

We performed a Monte Carlo optimization to obtain a robust fit for the radio light curves of SN\,2010P. 
We generated 10,000 random flux densities for each observing epoch with a detection, assuming that these 
follow a normal distribution with mean $S_{\nu}$ and standard deviation $\sigma$ (from Table \ref{tab:fluxcorr}). 
The random flux densities were restricted to positive values within the interval $-2\sigma$ to $+2\sigma$, thus 
covering the $95.4$ per cent of each epoch flux density distribution. 

\begin{figure}
\centering
\includegraphics[scale=0.655]{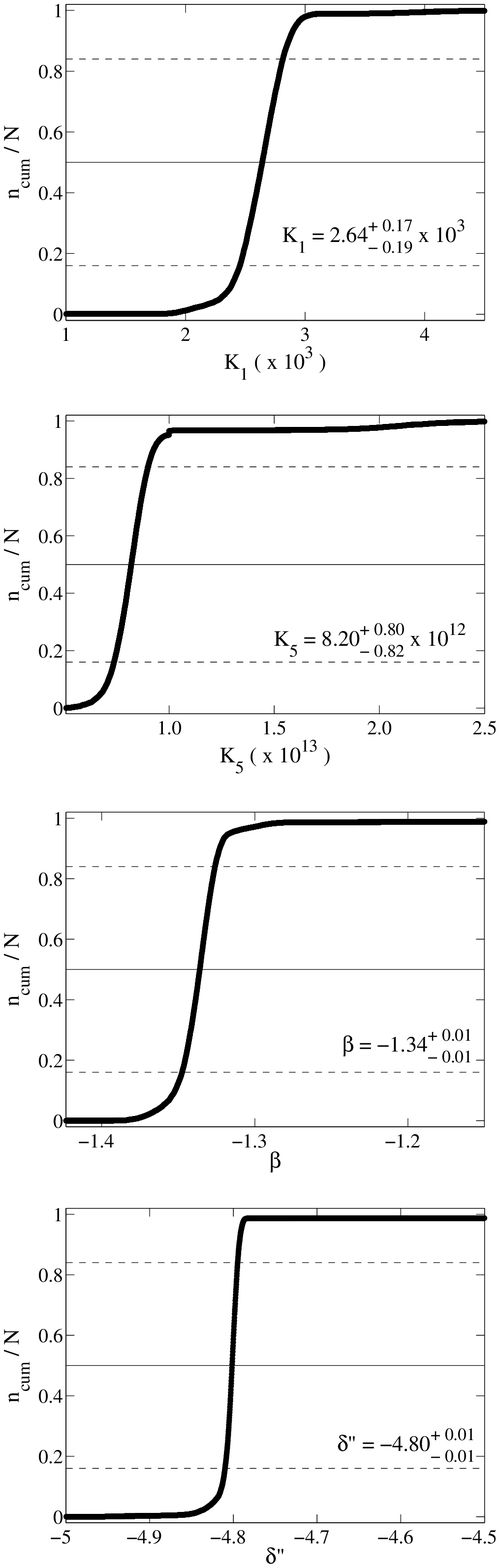}
\caption{Normalised cumulative distribution of the fitting parameters in the pure SSA model. The solid line represents the mean and
the dashed lines represent $-1\sigma$ and $+1\sigma$ of the distribution, repectively.} 
\label{fig:param_dist_SSA}
\end{figure}

\begin{figure}
\centering
\includegraphics[scale=0.655]{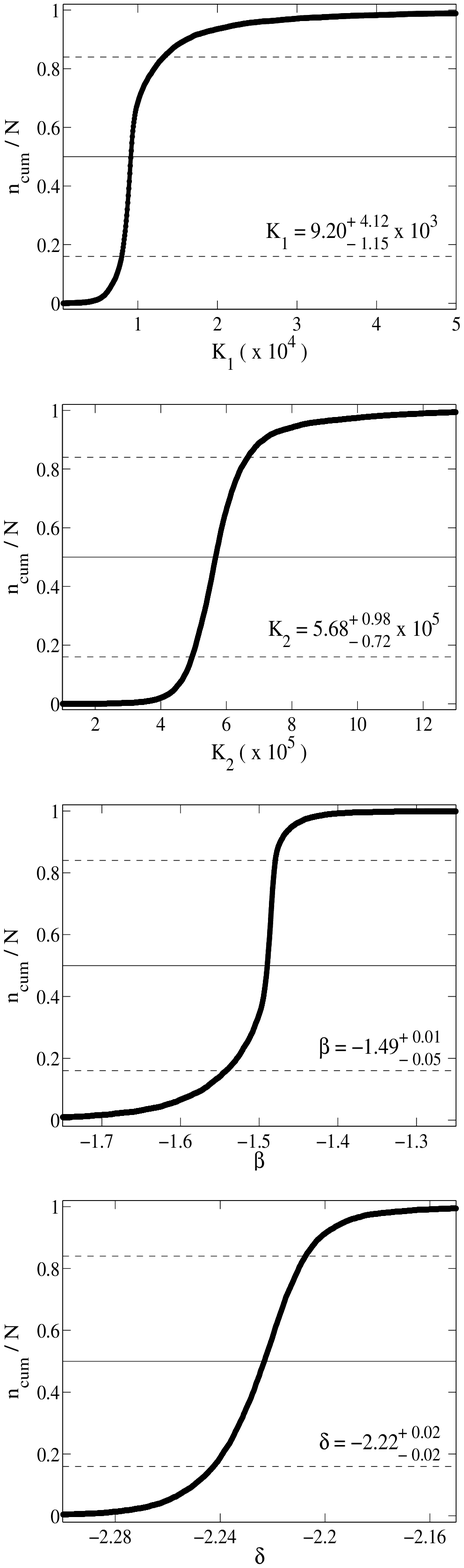}
\caption{Normalised cumulative distribution of the fitting parameters in the pure FFA model. Solid and dashed lines
are as described in Figure \ref{fig:param_dist_SSA}.} 
\label{fig:param_dist}
\end{figure}

We then created 10,000 sets (each composed of one flux value per epoch) to which we applied a non-linear 
least-squares regression method within MATLAB (from MathWorks). This process yielded a normal distribution 
for $K_1$, $K_5$, $\beta$ and $\delta^{\prime\prime}$ in the pure SSA model as shown in Figure 
\ref{fig:param_dist_SSA}, and for $K_1$, $K_2$, $\beta$ and $\delta$ in the pure FFA model as shown in 
Figure \ref{fig:param_dist}. For each distribution, i.e., for each parameter, the mean value is determined by 
the intersection of the solid line at $0.5$ with the curve, and the intersection with the dashed lines at $0.16$ 
and $0.84$ are the $-1\sigma$ and $+1\sigma$ values, respectively.

%%%%%%%%%%%%%%%%%%%%%%%%%%%%%%%%%%%%%%%%%%%%%%%%%%%%%%%%%%%%%%%%%%%%%%%%%%%%%%%%%%
%%%%%%%%%%%%%%%%%%%%%%%%%%%%%%%%%%%%%%%%%%%%%%%%%%%%%%%%%%%%%%%%%%%%%%%%%%%%% 
%%%%%%%%%%%%%%%%%%%%%%%%%%%%%%%%%%%%%%%%%%%%%%%%%%%%%%%%%%%%%%%%%%%%%%%%%%%%%%%%%%
%%%%%%%%%%%%%%%%%%%%%%%%%%%%%%%%%%%%%%%%%%%%%%%%%%%%%%%%%%%%%%%%%%%%%%%%%%%%% 

\subsection{Pure SSA fit parameters}\label{ap:mcexpt_SSA} 

The cumulative distributions for $K_1$, $K_5$, $\beta$ and $\delta^{\prime\prime}$, whilst consistent
with a Gaussian distribution, show some departures from the expected behaviour, as they reach a value of 
unity very abruptly. We note that the high value of $K_5$ implies that the SSA absorption is very important;
however, the high negative value of $\delta^{\prime\prime}$ indicates that the absorption decreases quite 
quickly with time, and tells us that SSA is more important at early stages in the evolution SN\,2010P. Thus, 
the poor sampling we have on the optically thick part of the radio light curves, inevitably results in a poor 
SSA fit.

\subsection{Pure FFA fit  parameters}\label{ap:mcexpt_FFA}

Parameter $K_1$ is the least symmetric in its cumulative distribution, however, it still resembles a Gaussian 
shape. The available number of epochs with detections at each frequency is not sufficient to constrain further 
parameters such as $K_3$ and $K_4$, and we therefore assumed them to be negligible. In Table \ref{tab:K3K4} we 
compare the values for $K_3$ and $K_4$ fitted for known Type II SNe and SN\,2010P. In most cases the term $K_4$ 
is absent, meaning that a putative distant ionized gas is not affecting the expansion of the SN ejecta. Thus, 
considering $K4 \sim 0$ for SN\,2010P is a fair assumption.

\begin{figure}
\centering
\includegraphics[width=84mm]{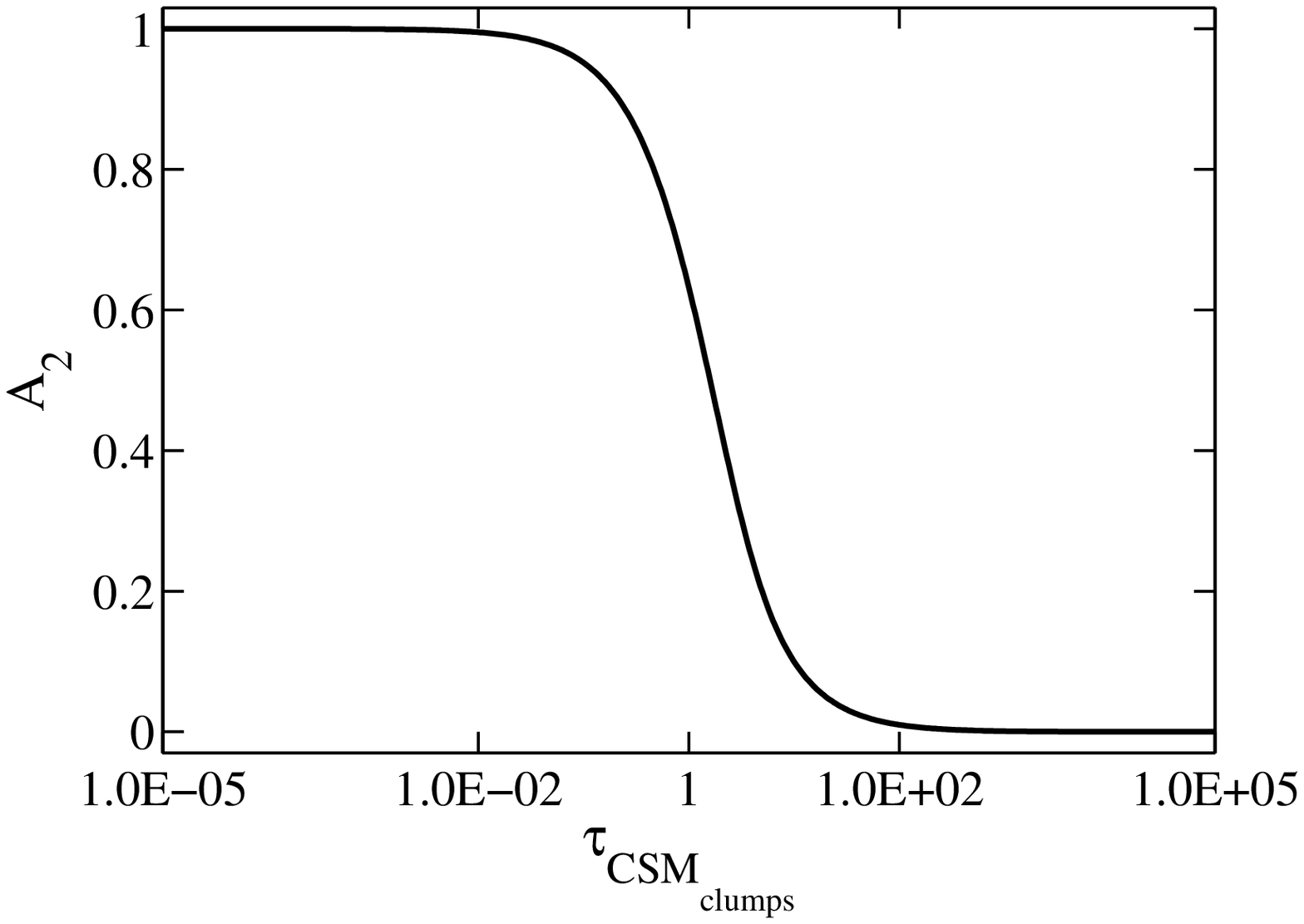}
\caption{Behaviour of the absorption by a clumpy medium.} \label{fig:tauclumps}
\end{figure}

In the case of $K_3$, we note however that when included in the fit of other SNe, it takes significantly high 
values. To explore its possible influence on the fit, we investigate the $A_2$ term in Equation \ref{eq:paramweiler}. 
$A_2$ takes values between 0 and 1, and for it to have some effect on the light-curve fitting requires that 
$0.01\la\tau_{\rmn{CSM}_{\rmn{clumps}}}\la100$ (see Figure \ref{fig:tauclumps}). This implies that a meaningful 
$K_3$ would take the following values at each of the frequencies we present data for: $6\e6 \la K_3 \la 2\ee12$, 
$ 6\e7\la K_3 \la 2\ee13$, $ 2\e8\la K_3 \la 5\ee13$ and $ 3\e9\la K_3 \la 7\ee14$ at 5.0 and 8.5\,GHz,
respectively. Therefore, $K_3 < 6\e6$ would have the same effect as $K_3 \sim 0$ at all the frequencies 
considered here. We cannot set strong constraints on $K_3$ owing to our sparse sampling at each frequency, and we 
note that the presence of a clumpy absorbing medium is not needed to reproduce the spectral behaviour of SN\,2010P. 
Thus, we assume $K_3 \sim 0$. 

%%%%%%%%%%%%%%%%%%%%%%%%%%%%%%%%%%%%%%%%%%%%%%%%%%%%%%%%%%%%%%%%%%%%%%%%%%%%%

\begin{table}
\centering
\caption{\protect{Absorption terms $K_3$ and $K_4$ from radio light-curve models of SN\,2010P and known Type II SNe.}} 
\label{tab:K3K4}
\begin{tabular}{ccc} \hline
SN & $K_3$ & $K_4$  \\
\hline		
2010P (IIb)$^a$  & 0.0              &  0.0       \\				     
1993J (IIb)$^b$  & $4.6\e5$         &  0.0       \\
2001gd (IIb)$^c$ &  -               &  -         \\
2001ig (IIb)$^d$ & $1.47\e5$        &  0.0       \\
2011hs (IIb)$^e$ & $1.9\e5$         &  0.0       \\
1970G (IIL)$^f$  & -                & -          \\
1979C (IIL)$^f$  & -                & -          \\
1980K (IIL)$^f$  & -                & -          \\
1978K (IIn)$^g$  & (0.03-30)$\ee11$ & $9\ee-3$   \\
1986J (IIn)$^f$  & $3.06\e9$        &  -         \\
1988Z (IIn)$^f$  & $5.39\e8$        &  0.0       \\
2000ft (IIn)$^h$ & -                & $\ga 0.17$ \\
    \hline
   \end{tabular} 
\begin{flushleft}
\vspace{-1mm}
The references are those from Table \ref{tab:SNcomp}.
\end{flushleft}
\end{table}
%%%%%%%%%%%%%%%%%%%%%%%%%%%%%%%%%%%%%%%%%%%%%%%%%%%%%%%%%%%%%%%%%%%%%%%%%%%%%%%%%%
%%%%%%%%%%%%%%%%%%%%%%%%%%%%%%%%%%%%%%%%%%%%%%%%%%%%%%%%%%%%%%%%%%%%%%%%%%%%%%%%%%
%%%%%%%%%%%%%%%%%%%%%%%%%%%%%%%%%%%%%%%%%%%%%%%%%%%%%%%%%%%%%%%%%%%%%%%%%%%%%

\label{lastpage}

\end{document}